# PRICING DERIVATIVES IN HERMITE MARKETS

**Stoyan V. Stoyanov**
*College of Business*
*Stony Brook University*
Email: stoyan.stoyanov@stonybrook.edu

**Svetlozar T. Rachev**
*Texas Tech University and GlimmAnalytics*
Email: zarirachev1951@gmail.com

**Stefan Mittnik**
*Ludwig Maximilians University Munich*
Email: mittnik@gmx.de

**Frank J. Fabozzi**
*EDHEC Business School*
Email: fabozzi321@aol.com

**Abstract:** We present a new framework for Hermite fractional financial markets, generalizing the fractional Brownian motion and fractional Rosenblatt markets. Considering pure and mixed Hermite markets, we introduce a strategy-specific arbitrage tax on the rate of transaction volume acceleration of the hedging portfolio as the prices of risky assets change, allowing us to transform Hermite markets with arbitrage opportunities to markets with no arbitrage opportunities within the class of Markov trading strategies.

**Keywords**: Hermite processes; fractional Brownian motion; Rosenblatt processes; perpetual derivatives; no-arbitrage; arbitrage tax

**JEL**: G13, G11, G19

## 1. INTRODUCTION

In this paper, we introduce a novel method for pricing derivatives in fractional markets. All existing fractional market models[1] assume that the riskless asset has the same dynamics as in the classical option pricing models developed by Black and Scholes (1973) and Merton (1973), hereafter referred to as the BSM model. But that assumption leads to the existence of arbitrage trading strategies (see Rogers (1997) and Shiryayev (1998)). Applying Wick integration, Hu and Øksendal (2003) show that a fractional Brownian motion (FBM) market has no arbitrage opportunities. However, Wick integration has no economic intuition (see, for example, Björk and Hult (2005)) and the corresponding replicating strategies are very restrictive.[2]

We propose a different approach in this paper. We first postulate that in arbitrage-free complete fractional markets, exhibiting long-range dependence (LRD), the riskless asset, if publicly traded,[3] should be a perpetual derivative of the risky assets. Second, we generalize FBM by using Hermite processes which are also inherently clairvoyant and admit arbitrage opportunities. However, instead of using the approach of Cheredito (2003) and ruling out arbitrage opportunities by limiting the set of admissible strategies to only those that do not trade in arbitrarily small time intervals, we suggest the introduction of a strategy-specific arbitrage tax without imposing a severe restriction on the set of admissible strategies. As a result, Hermite markets become free of arbitrage opportunities when financing strategies are restricted to Markov strategies[4] with a positive, strategy-specific arbitrage tax —regardless of how small the tax is. We investigate the effect of such taxes in the context of the classical case of BSM diffusion markets and show that, as far as pricing is concerned, if the market is arbitrage free, the introduction of an arbitrage tax in hedging a derivative is equivalent to an increase of the underlying risky asset's volatility. We also provide an example of a BSM market with arbitrage opportunities which, after the introduction of an arbitrage tax becomes arbitrage free—again, regardless of how small the tax is.

---

[1] See, for example, Sottinen (2001), Bender, Sottinen, and Valkeila (2007), Mishura (2008), and Rostek (2009).

[2] In the FBM-market setup of Elliot and Van der Hoek (2003), the simple buy-and-hold strategy is not self-financing (which is indeed discouraging).

[3] In this paper, we do not question the existence of a riskless asset. This has been subject of considerable debate recently with Fisher (2013) arguing: "The idea of risk-free sovereign bonds is best thought of as an oxymoron or as an anomaly of recent history. It is not a useful, necessary or an enduring feature of the financial landscape."

The paper is organized as follows. In Section 2 we study pure Hermite markets, demonstrating that under the new arbitrage tax the markets become arbitrage free. Our choice for using Hermite markets to model a general fractional market is motivated by the flexibility[5] of Hermite motion as a self-similar process with stationary increments having Gaussian or heavy-tailed distributions; FBM is a special case of Hermite motion. In Section 3 we extend our framework to mixed Hermite markets, that is, markets with two independent drivers: a Brownian motion and a Hermite motion. Concluding remarks are given in Section 4. In Appendix A1, we summarize all basic facts on Hermite motions. Essential proofs are presented in Appendix A2. Proofs involving standard no-arbitrage derivative valuation arguments[6] are omitted.

[4] A self-financing strategy is a Markov strategy if it is a smooth function of the price processes defining the dynamics of the market securities.

## 2. THE PURE HERMITE MARKET MODEL

We suggest Hermite motion as a building block of the stochastic driver of a price process for two reasons. Firstly, it is a self-similar process which contains FBM as a special case and can account for the long-range dependence phenomenon exhibited by real-world by financial markets. Secondly, apart from the special case of FBM, Hermite motion is not a Gaussian process and can accommodate fat-tail phenomena present in real-world financial markets.

To set up our framework, we start with the definition of our market model and then develop the notion of arbitrage tax which leads to an arbitrage-free market.Let $\mathcal{H}^{(H,\mathscr{k}(i))}(t), t \geq 0, \mathscr{k}(i) \in \mathbb{N}, H \in \left(\frac{1}{2}, 1\right), i = 1, \dots, I,$ be $I$ Hermite motions given by

(2.1) $\mathcal{H}^{(H,\mathscr{k}(i))}(t) = C^{(H,\mathscr{k}(i))} \int_{\mathcal{D}^{\mathscr{k}}} K_t^{(H,\mathscr{k}(i))}\left(v^{(1)}, \dots, v^{(\mathscr{k}(i))}\right) dB^{(i)}\left(v^{(1)}\right) \dots dB^{(i)}\left(v^{(\mathscr{k}(i))}\right), t \geq 0,$

and generated by independent two-sided Brownian motions (BMs), $B^{(i)}(v), v \in R,$ defined on $(\Omega, \mathcal{F}, \{\mathcal{F}_t\}_{t \geq 0}, \mathbb{P})$, where

$$K_t^{(H,\mathscr{k}(i))}(\mathbb{v}) := \int_0^t \left[ \prod_{j=1}^{\mathscr{k}(i)} \left(s - v^{(j)}\right)_+^{\frac{H-1}{\mathscr{k}(i)} - \frac{1}{2}} \right] ds$$

and $C^{(H,\mathscr{k}(i))}$ is a normalizing constant. See Appendix A.1 for additional details and properties of Hermite motion.

---

[6] See, for example, Duffie (2001, Chapter 6).

Let $\mathbb{a} = \left(a^{(1)}, \dots, a^{(I)}\right) \in (0,\infty)^I$, $\sum_{i=1}^{I} {a^{(i)}}^2 = 1$, and set

(2.2) $\mathcal{H}(t) \coloneqq \mathcal{H}^{(\mathbb{a})}(t) := \sum_{i=1}^{I} a^{(i)} \mathcal{H}^{(H,\mathscr{k}(i))}(t), t \geq 0.$

Note that process $\mathcal{H}(t), t \geq 0$, referred to as a *mixed Hermite process*, will play the role of market driver in all fractal market models used throughout this paper.

2.1 Arbitrage in the Pure Hermite Market

Designating the risky asset in the *pure Hermite market* by $\mathbb{S}^{(PH)}$, we assume that $\mathbb{S}^{(PH)}$dynamics follows the *geometric mixed Hermite process*

(2.3) $S^{(PH)}(t) = S^{(PH)}(0) \exp\{\mu(t) + \sigma(t)\mathcal{H}(t)\}, t \geq 0, S^{(PH)}(0) > 0,$

where $\sigma(t) > 0, t \geq 0$, $\mu \in C^{(1)}[[0,\infty)]$,[7] $\sigma \in C^{(1)}[[0,\infty)]$. Throughout the literature,[8] it has been assumed that the riskless asset (bond), typically denoted as $\mathcal{M}$, has price dynamics

---

[7] That is, $\mu: [0,\infty) \to R$, has a continuous first derivative.

[8] See the surveys by Mishura (2008), Biagini et al. (2008), and Rostek (2009). As we shall see from Propositions 1 and 2, the market model with primary assets $\mathbb{S}^{(PH)}$ and $\mathcal{M}$ has an economic meaning only if arbitrage taxes are included in the hedging portfolios.

(2.4) $\beta(t)=\beta(0)e^{rt}, t\geq 0, r\in R, \beta(0)>0, r\in R.$

While in the BSM-market, (2.4) is the correct choice for the dynamics of the riskless asset $\mathcal{M}$,[9] in fractal markets this is not always the case.[10] To demonstrate that, in general, $\mathcal{M}$ should not be used[11] as the traded security paired with risky asset $\mathbb{S}^{(PH)}$, let us replicate the approach for a market with two risky securities formulated by Black (1972),[12] and determine the dynamics of the riskless asset (designated as $\mathcal{M}$ ($\mathcal{H}$)). Consider the risky assets $\mathbb{S}^{(i)}, i = 1, \dots, N$, with price processes

(2.5) $$S^{(PH,i)}(t) = S^{(PH,i)}(0) \exp\{\mu^{(i)}(t) + \sigma^{(i)}(t)\mathcal{H}(t)\}, t \geq 0,$$

---

[9] See the seminal work by Black (1972); Rachev and Fabozzi (2016) provide a review of the topic.

[10] Market models with primary assets $\mathbb{S}^{(PH)}$ and $\mathcal{M}$ have been proposed in the literature and wrongly claimed to be option pricing models; see Necula (2002), Sottinen andValkeila, (2003), Mishura (2008, Section 5.2), Xiao et al. (2010), and Shokrollahi and Kılıçman (2016), where the European option pricing formula in the market $\left(\mathbb{S}^{(PH)},\mathcal{M}\right)$ is the particular FBM $\mathcal{H}(t)=B^{(H)}(t), \mu(t)=at-\frac{1}{2}\sigma^2t^{2H}, t\geq 0, a\in R, \sigma(t)=\sigma>0$. As we shall show in Proposition 1, all those models are misleading because they are not arbitrage-free within the class of Markov trading strategies (introduced in Mishura (2008)). Non-Markov trading strategies for fractal markets with $\mathbb{S}^{(PH)}$ as the risky asset do not have an economic meaning. See, for example, Mishura (2008) and Rostek (2009) for further discussions on this issue.

[11]This is under the assumption that the model is frictionless, i.e., no arbitrage taxes are included. See Proposition 1.

[12] Black's approach is more general, considering $N+1$ securities with prices following geometric Brownian motions (GBMs) with $N$- independent BMs. We adopt Black's approach, using two securities to better illustrate the results. The extension to $N+1$ securities with $N$-

$S^{(PH,i)}(0) = x^{(i,0)} > 0, i = 1, \dots, N,$ where $\mu^{(i)}, \sigma^{(i)} \in C^{(1)}[[0, \infty)], i = 1, \dots, N,$ and define $\mathcal{S}^{(PH)}(t) := \left(S^{(PH,1)}(t), \dots, S^{(PH,N)}(t)\right), t \geq 0.$

Given the market $\mathfrak{S}_H := \left(\mathbb{S}_H^{(1)}, \dots, \mathbb{S}_H^{(N)}\right)$, consider the class of all **Markov self-financing strategies**, $a\left(\mathcal{S}^{(PH)}(t)\right) := \left(a^{(1)}\left(\mathcal{S}^{(PH)}(t)\right), \dots, a^{(N)}\left(\mathcal{S}^{(PH)}(t)\right)\right), \mathrm{t} \geq 0,$ that is, the corresponding self-financing portfolios $\mathrm{P}^{(\mathrm{M})}(\mathrm{t}) = \mathcal{P}\left(\mathcal{S}^{(PH)}(t)\right), \mathrm{t} \geq 0,$ have the representation:[13]

---

independent BMs as market drivers can be studied in a similar fashion as in Rachev and Fabozzi (2016).

[13] Note that $\int_0^t a^{(i)}\left(\mathcal{S}^{(PH)}(t)\right) d^{(S)} S^{(PH,i)}(t)$ is understood as a pathwise integral, that is, as a limit of a sequence of gain processes of buy-and-hold-strategies,

$\int_0^t a^{(i)}\left(\mathcal{S}^{(PH)}(t)\right) d^{(S)} S^{(PH,i)}(t) =$

$lim_{n\uparrow\infty\left(t^{(k)}=\frac{k}{n}t, k=0,\dots,n\right)} \sum_{k=0}^{n-1} a^{(i)}\left(\mathcal{S}^{(PH)}\left(t^{(k)}\right)\right)\left(\mathcal{S}^{(PH)}\left(t^{(k+1)}\right) - \mathcal{S}^{(PH)}\left(t^{(k)}\right)\right).$

This is the main advantage of choosing pathwise integration versus the alternative fractal integrations, such as Wick integration. See, for example, Mishura (2008, pp. 323-324) and Rostek (2009, pp. 65-66).

(2.6) $\mathcal{P}\left(\mathcal{S}^{(PH)}(t)\right) = \sum_{i=1}^{N} a^{(i)}\left(\mathcal{S}^{(PH)}(t)\right) S^{(PH,i)}(t)$

$$= \mathcal{P}\left(\mathcal{S}^{(PH)}(0)\right) + \sum_{j=1}^{N} \int_0^t a^{(i)}\left(\mathcal{S}^{(PH)}(s)\right) d^{(S)} S^{(PH,i)}(s) \, .$$

In (2.6), the functions $a^{(j)}(\mathbb{x}), \mathrm{j} = 1, \dots, \mathrm{N}$, and

(2.7) $\mathcal{P}(\mathbb{x}) = \sum_{i=1}^{N} x^{(i)} \, a^{(i)}(\mathbb{x}), \ \mathbb{x} = \left(x^{(1)}, \dots, x^{(N)}\right) \in R_+^N \coloneqq (0, \infty)^N$

have continuous first-order derivatives.[14] In the next proposition, without loss of generality, we assume that $S^{(PH,i)}(0) = 1, i = 1, \dots, N$.

*PROPOSITION 1: Within the class of Markov self-financing strategies, the corresponding portfolios, satisfying (2.6) and (2.7), will also satisfy the partial differential equation (PDE)*

(2.8) $\sum_{j=1}^{N} \frac{\partial \mathcal{P}(\mathbb{x})}{\partial x^{(j)}} - \mathcal{P}(\mathbb{x}) = 0, \mathbb{x} \in R_+^N.$

*In particular,* for every $c = \left(c^{(i,j)} > 0, i, j = 1, \dots, N\right)$, $\mathcal{P}^{(c)}(\mathbb{x}) = \sum_{i=1}^{N} \sum_{j=1}^{N} c^{(i,j)} \left(\sqrt{x^{(i)}} - \sqrt{x^{(j)}}\right)^2$ *satisfies (2.8), and* $\mathcal{P}^{(c)}\left(\mathcal{S}^{(PH)}(t)\right), t \geq 0,$ *is an arbitrage portfolio.*

*Proof of Proposition 1:* See Appendix A2.1

---

[14] In short, $a^{(j)}, \mathcal{P} \in \mathrm{C}^{(I)}[R_+^N], \ I = (1, \dots, 1) \in R_+^N.$

Proposition 1 provides many examples of arbitrage Markov self-financing strategies, such as $\mathcal{P}^{(c)}(\mathbb{x}) = \left(\sum_{i=1}^{N} a^{(i)}\sqrt{x^{(i)}}\right)^2$, $a^{(i)} \in R$, $\sum_{i=1}^{N} a^{(i)} = 0$. Shiryayev (1998)[15] introduced a similar example of an arbitrage strategy for the special case of $N = 2, \sigma^{(1)} = 0, \sigma^{(2)}(t) = \sigma^{(2)} > 0$, $\mu^{(i)}(t) = r, r = 1,2$, and $\mathcal{H}(t) = B^H(t), t \geq 0$.

Note that, according to Proposition 1, every perpetual derivative $\mathfrak{G}^{(\rho)}, \rho = (\rho(1), \dots \rho(N)) \in R^N$, with price process $\mathcal{G}^{(\rho)}\left(\mathcal{S}^{(PH)}(t)\right) = \prod_{j=1}^{N} S^{(PH,j)}(t)^{\rho(j)}, t \geq 0$, can be replicated by a Markov strategy, if and only if $\sum_{j=1}^{N} \rho(j) = 1$. Thus, the riskless security $\mathfrak{M}^{(r)}$ with price dynamics $\mathfrak{m}(t) = e^{\mathfrak{R}(t)}, t \geq 0$, with riskless rate $r(t) = \frac{\partial \mathfrak{R}(t)}{\partial t}, t \geq 0$, exists, if $\mathfrak{R}(t), t \geq 0$ has the representation $\mathfrak{R}(t) = \sum_{j=1}^{N} \breve{\rho}(j)\, \mu^{(i)}(t), t \geq 0$, where $\breve{\rho}(j) \in R, j = 1, \dots, N, \sum_{j=1}^{N} \breve{\rho}(j) = 1$, and $\sum_{j=1}^{N} \breve{\rho}(j)\, \sigma^{(i)}(t) = 0$. As a result, we have created a synthetic riskless asset $\mathfrak{M}^{(r)}$, as a perpetual derivative in the market $\mathfrak{S}_H := \left(\mathbb{S}_H^{(1)}, \dots, \mathbb{S}_H^{(N)}\right)$.

Unfortunately, the extended market $\left(\mathfrak{M}^{(r)}, \mathfrak{S}_H\right)$ is not arbitrage free as Proposition 1 claims. Nevertheless, our methodology of seeking the appropriate dynamics of the riskless asset within the market of risky assets will only be explored when seeking Hermite markets with no arbitrage opportunities.

---

[15] Consider a pure Hermite market with zero interest rate: $S(t) = e^{\mathcal{H}(t)}, \beta(t) = 1, t \geq 0$. Form a strategy, $a(t) = 2S(t) - 2, b(t) = 1 - S(t)^2$, generating a self-financing portfolio $P^{(H,k)}(t) := a(t)S(t) + b(t)\beta(t) = (S(t) - 1)^2$. Clearly, $P^{(H,k)}(0) = 0$, while $P^{(H,k)}(t) > 0$, $\mathbb{P}$-a.s. Shiryayev (1998) provided this example for the case of FBM, $B^H = \mathcal{H}^{(H,1)}$. Note that $a(t), b(t), t \geq 0$, are smooth functions of the stock price $S(t)$ only; that is, they are *Markov self-financing strategies.* See also Mishura (2008, p. 304) and Rostek (2009, p. 58).

2.2 No-Arbitrage in the Pure Hermite Market with Arbitrage Taxes

Does the fact that the market model $\mathfrak{S}_H \coloneqq \left(\mathbb{S}_H^{(1)}, \dots, \mathbb{S}_H^{(N)}\right)$ exhibits arbitrage opportunities mean we should abandon its use? If all market participants are using (2.3) and (2.4) as the market model, what will be the "fair price" of a perpetual derivative? It turns out that the introduction of strategy-specific arbitrage taxes in the market model $\mathfrak{S}_H \coloneqq \left(\mathbb{S}_H^{(1)}, \dots, \mathbb{S}_H^{(N)}\right)$—no matter how small—will offset fractal market price predictive/clairvoyant features. Indeed, we have shown that there exist arbitrage Markov strategies in a Hermite market, and thus the trading strategies should be subject to some cost attributable to regulatory restrictions, market frictions, moral hazard penalty, and the like.

In fact, the suggested arbitrage tax can be related to the idea of imposing hedging transaction costs when the stock price is a geometric fractional Brownian motion, as was investigated by Guasoni, Rásonyi, and Schachermayer (2008), Rostek (2009), Wang (2010), and Gu, Liang, and Zhang (2012). In the discrete binary setting, transaction costs are imposed by trading on selected nodes of the fractional binary trees, see Rostek (2009). In continuous time, the transaction costs are proportional to the value of the transactions in the underlying stock; see, for example, Biagini et al. (2008) and Nteumagné, Pindza, and Maré (2014). Our approach is different in that we assume that the arbitrage tax is determined by the rate of transaction volume acceleration of the hedging portfolio which we refer to as the *velocity of hedging*.

In view of this discussion, let us suppose that a Markov trading strategy in a Hermite market comes at a cost due to a strategy-specific arbitrage tax. We require that the portfolio dynamics

(2.9) $\mathcal{P}^{(\mathrm{C})}\left(\mathcal{S}^{(PH)}(t)\right) = \sum_{j=1}^{N} a^{(C,j)}\left(\mathcal{S}^{(PH)}(t)\right) \mathcal{S}^{(PH,j)}(t)$

involve Markov strategies $\left(a^{(C,1)}\left(\mathcal{S}^{(PH)}(s)\right), \dots, a^{(C,N)}\left(\mathcal{S}^{(PH)}(s)\right)\right)$ with embedded running cost functional

(2.10) $\mathcal{P}^{(\mathrm{C})}\left(\mathcal{S}^{(PH)}(t)\right) - \mathcal{P}^{(\mathrm{C})}\left(\mathcal{S}^{(PH)}(0)\right) =$

$$= \sum_{j=1}^{N} \int_0^t a^{(C,j)}\left(\mathcal{S}^{(PH)}(s)\right) d^{(S)} \mathcal{S}^{(PH,j)}(s) - C\left(\mathcal{S}^{(PH)}(t)\right), t \geq 0,$$

where $C\left(\mathcal{S}^{(PH)}(t)\right), t \geq 0$ is the running cost function.

An immediate issue is the representation of $C\left(\mathcal{S}^{(PH)}(t)\right)$. Clearly, $C\left(\mathcal{S}^{(PH)}(t)\right)$ should depend on the dynamics of $\mathcal{P}\left(\mathcal{S}^{(PH)}(t)\right)$; and we demonstrate that it is natural to have it depend on the velocity of $\mathcal{P}\left(\mathcal{S}^{(PH)}(t)\right), t \geq 0$. In view of that, we set

(2.11) $C\left(\mathcal{S}^{(PH)}(t)\right) = \frac{1}{2}\sum_{j=1}^{N} \int_0^t {c^{(j)}}^2 \frac{\partial^2 \mathcal{P}^{(\mathrm{C})}\left(\mathcal{S}^{(PH)}(s)\right)}{\partial {x^{(j)}}^2} \mathcal{S}^{(PH,j)}(s) d^{(S)} \mathcal{S}^{(PH,j)}(s), t \geq 0,$

for some $c^{(j)} > 0$.[16] The integrand involves the second derivative of the portfolio value with respect to the corresponding state variable. This implies that portfolios linear in the state variables (buy-and-hold portfolios) have a zero arbitrage tax. A positive arbitrage tax appears only if the quantities held of each risky asset depend on the price of the corresponding asset. In the following, we provide a simple example which allows us to compute the arbitrage tax explicitly. The strategy consists of a risky asset and a bond which we assume has a constant price of one. This setting is similar to that in Section 5 of Föllmer and Schied (2013).

[16] We call $c^{(j)} > 0$ the tax-intensity of asset $\mathbb{S}_H^{(j)}$.

Consider the following self-financing strategy: $V_t = f'\big(S(t)\big)S(t) + \eta_t$ invested in $S(t)$ and a bond with a constant price of one in which the function $f$ is sufficiently smooth. The strategy is self-financing,

$$V_t = V_0 + \int_0^t f'\big(S(s)\big)dS(s) = V_0 + f\big(S(t)\big),$$

where the integral is path-wise and we assume the quadratic variation of $S(t)$ vanishes, a property that holds for Hermite motions. The investment in the bond equals $\eta_t = f\big(S(t)\big) - f'\big(S(t)\big)S(t)$. We further assume $V_0 = 0$ and that the above is an arbitrage strategy as per the definition in Cheridito (2003); that is, $f\big(S(t)\big)$ is $[0, \infty]$-valued and $P\{f\big(S(t)\big) > 0\} > 0$.

Next, we calculate the arbitrage tax for this strategy. The portfolio $\mathcal{P}(\mathbb{x}) = \mathcal{P}(x, y)$ in equation (2.7) equals

$$\mathcal{P}(x, y) = f'(x)x + (f(x) - f'(x)x)y$$

and the corresponding second-order derivatives are

$$\frac{\partial^2 \mathcal{P}(x, y)}{\partial x^2} = xf'''(x) + 2f''(x) - \big(f''(x) + xf'''(x)\big)y$$

$$\frac{\partial^2 \mathcal{P}(x, y)}{\partial y^2} = 0.$$

Then, the second derivative in the expression for the arbitrage tax in (2.11) is $f''\big(S(s)\big)$.

The cost function can be expressed explicitly by,

$$C_t = \frac{c^2}{2}\int_0^t \frac{\partial^2 \mathcal{P}(S(s),1)}{\partial x^2} S(s)dS(s)$$

$$= \frac{c^2}{2}\int_0^t f''\big(S(s)\big)S(s)dS(s)$$

$$= \frac{c^2}{2}\Big(f'\big(S(t)\big)S(t) - f\big(S(t)\big) - f'\big(S(0)\big)S(0)\Big),$$

after noticing that $f'(x)x - f(x)$ is the primitive function of $xf''(x)$ and that $f\big(S(0)\big) = 0$ by construction. As a result, the portfolio value process net of tax is $V_t - C_t = f\big(S(t)\big) - \frac{c^2}{2}\Big(f'^{(S(t))}S(t) - f\big(S(t)\big) - f'^{(S(0))}S(0)\Big),$ which is no longer guaranteed to be positive with probability one. The instantaneous tax controls the probability that the value process net of tax is positive,

$$P\{V_t - C_t > 0\} = P\left\{f\big(S(t)\big) > \frac{c^2}{2}\Big(f'\big(S(t)\big)S(t) - f\big(S(t)\big) - f'\big(S(0)\big)S(0)\Big)\right\}$$

As a consequence, the portfolio value net of tax is no longer a free lunch with vanishing risks as defined in Ceridito (2003).

The example in Section 5 in Föllmer and Schied (2013) becomes a special case with $f(x) = \big(x - S(0)\big)$.[2] The above probability then becomes $P\left\{\big(S(t) - S(0)\big)^2 > \frac{c^2}{2}\big(S^2(t) - S^2(0)\big)\right\} = P\left\{S(t) > \frac{1+\frac{c^2}{2}}{1-\frac{c^2}{2}}S(0) | S(t) > S(0)\right\}P\{S(t) > S(0)\} + P\{S(t) < S(0)\}$.

*PROPOSITION 2: Suppose the* $\mathfrak{S}_H \coloneqq \left(\mathbb{S}_H^{(1)}, \dots, \mathbb{S}_H^{(N)}\right)$ *is a market consisting only of risky securities with price processes*

(2.12) $S^{(PH,j)}(t) = \exp\{\mu^{(j)} t + \sigma^{(j)} \mathcal{H}(t)\}, t \geq 0, \mu^{(j)} \in R, \sigma^{(j)} > 0, j = 1, .., N.$

*Within the class of Markov strategies with the cost constraints given by (2.10) and (2.11), the corresponding portfolios* $\mathcal{P}^{(C)}\left(\mathcal{S}^{(PH)}(t)\right), t \geq 0$*, defined by (2.9), satisfy*

(2.13) $$\sum_{j=1}^{N} \frac{\partial \mathcal{P}^{(C)}(\mathbb{x})}{\partial x^{(j)}} x^{(j)} - \mathcal{P}^{(C)}(\mathbb{x}) + \sum_{j=1}^{N} \frac{1}{2} c^{(j)^2} \frac{\partial^2 \mathcal{P}^{(C)}(\mathbb{x})}{\partial x^{(j)^2}} x^{(j)^2} = 0 \quad , \mathbb{x} \in R_+^N.$$

*Suppose that* $\sigma^{(j)} > 0, j = 1, .., N$ *satisfy* $\sum_{j=1}^{N} \sigma^{(i)} \phi^{(j)} = 0$ *for some* $\phi^{(j)} \in R$*, such that* $\sum_{j=1}^{N} \phi^{(j)} - 1 + \sum_{j=1}^{N} \frac{1}{2} c^{(j)^2} \phi^{(j)} \left(\phi^{(j)} - 1\right) = 0$. *Then,* $\mathfrak{S}_H \coloneqq \left(\mathbb{S}_H^{(1)}, \dots, \mathbb{S}_H^{(N)}\right)$ *determines a riskless asset* $\mathcal{M}$ *with price dynamics*

(2.14) $$\beta(t) = e^{rt}, t \geq 0,$$

*with riskless rate* $r = \sum_{j=1}^{N} \mu^{(j)} \phi^{(j)}$*. Consider the extended market*

$\mathfrak{M}^{(H)} = \left(\mathbb{S}_H^{(0,)}, \mathbb{S}_H^{(1)}, \dots, \mathbb{S}_H^{(N)}\right), \mathbb{S}_H^{(0)} = \mathcal{M}$ *and assume that (2.9), (2.10), and (2.11) hold. Then*

$\mathfrak{M}^{(H)}$ *is free of arbitrage within the class of Markov strategies with costs given by (2.9), (2.10), and (2.11).*

*Proof of Proposition 2:* See Appendix A2.2.

To illustrate the nature of the arbitrage tax we shall apply it to the classic BSM model. Let $\mathfrak{S} := \left(\mathbb{S}^{(1)}, \mathbb{S}^{(2)}\right)$ be a market of two risky securities with price processes $\mathcal{S}(t) = \left(S^{(1)}(t), S^{(N)}(t)\right)$ with

(2.15) $S^{(j)}(t) = \exp\left\{\left(\mu^{(j)} - \frac{1}{2}{\sigma^{(j)}}^2\right)t + \sigma^{(j)}W(t)\right\}, t \geq 0, \mu^{(j)} \in R, \sigma^{(1)} > \sigma^{(2)} > 0, j = 1,2,$

where $W(t), t \geq 0,$ is a standard BM, generating the stochastic basis $(\Omega, \mathcal{F}, \{\mathcal{F}_t\}_{t\geq 0}, \mathbb{P})$. Consider a perpetual derivative, designated as $\mathfrak{D}$, with a price process $\mathcal{G}(t) = g\left(\mathcal{S}(t)\right), t \geq 0$, where $g \in C^2[R_+^N]$[17]. Assume that trader ב holds a short position in the $\mathfrak{D}$-contract. Furthermore, ב's replicating portfolio $\mathcal{P}\left(\mathcal{S}(t)\right) = g\left(\mathcal{S}(t)\right), \mathrm{t} \geq 0,$

(2.16) $\mathcal{P}\left(\mathcal{S}(t)\right) = \sum_{j=1}^{2} a^{(j)}\left(\mathcal{S}(t)\right) S^{(j)}(t), t \geq 0,$

involves an arbitrage tax given by

(2.17) $\mathcal{P}(\mathrm{t})\left(\mathcal{S}(t)\right) - \mathcal{P}(\mathrm{t})\left(\mathcal{S}(0)\right) = \sum_{j=1}^{2} \int_0^t a^{(j)}\left(\mathcal{S}(s)\right) dS^{(j)}(s) - \hat{C}\left(\mathcal{S}(t)\right), t \geq 0,$

where

(2.18) $\hat{C}\left(\mathcal{S}(t)\right) = \frac{1}{2}\sum_{j=1}^{2} \int_0^t {\hat{c}^{(j)}}^2 \frac{\partial^2 \mathcal{P}\left(\mathcal{S}(s)\right)}{\partial {\mathrm{x}^{(j)}}^2} S^{(j)}(s) dS^{(j)}(s),\ t \geq 0,\ \hat{c}^{(j)} > 0,\ j = 1,2.$

Then, straightforward hedging arguments[18] lead to the following PDE for $g(\mathbb{x}), \mathbb{x} \in R_+^2$:

(2.19) $r\frac{\partial g(\mathbb{x})}{\partial x^{(1)}} x^{(1)} + r\frac{\partial g(\mathbb{x})}{\partial x^{(2)}} x^{(2)} - rg(\mathbb{x}) + \frac{1}{2}\frac{\partial^2 g(\mathbb{x})}{\partial {\mathrm{x}^{(1)}}^2} {x^{(1)}}^2 \left({\sigma^{(1)}}^2 + r{\hat{c}^{(1)}}^2\right) +$

[17] That is, $g(\mathbb{x}), \mathbb{x} \in R_+^2 = (0, \infty)^2$ has continuous derivatives of second order.

[18] See, for example, Duffie (2001, Chapter 6).

$$\frac{1}{2}\frac{\partial^2 g(\mathbb{x})}{\partial \mathrm{x}^{(2)^2}} x^{(2)^2}\left(\sigma^{(2)^2} + r\hat{c}^{(2)^2}\right) = 0,$$

where

(2.20) $$r \coloneqq \frac{\mu^{(2)}\sigma^{(1)} - \mu^{(1)}\sigma^{(2)}}{\sigma^{(1)} - \sigma^{(2)}}.$$

PDE (2.19) is the classical BSM-PDE for a derivative in the market with two risky assets and no riskless asset, where the instantaneous variances of the asset returns are increased by the tax intensities $r\hat{c}^{(j)^2}, j = 1,2$. In particular, if $\mathfrak{D}^{(a,b)}$ is a security with price process $g^{(a,b)}(\mathcal{S}(t)) = S^{(1)^a} S^{(2)^b}, a, b \in R$, then $g^{(a,b)}(\mathbb{x}), \mathbb{x} \in R_+^2$, satisfies (2.19) if and only,

$$ra + rb - r + \frac{1}{2}a(a-1)\left(\sigma^{(1)^2} + rc^{(1)^2}\right) + \frac{1}{2}b(b-1)\left(\sigma^{(2)^2} + rc^{(2)^2}\right) = 0.$$

Next, given a market with riskless asset $\mathcal{M} = \mathbb{S}^{(1)}$ and risky asset $\mathbb{S} = \mathbb{S}^{(2)}$, with price process

(2.21) $$\mathrm{S}(t) = \exp\left\{\left(\mu - \frac{1}{2}\sigma^2\right)t + \sigma W(t)\right\}, t \geq 0, \mu \in R, \sigma > 0,$$

let us assume that the cost-intensity for the riskless asset $\mathcal{M}$ is zero, while the cost-intensity for the risky asset, $\mathbb{S}$, is $\hat{c} > 0$. Consider a perpetual derivative $\mathfrak{D}^{(\mathcal{M},\mathbb{S})}$ with price process $\mathfrak{h}(t) = \hbar(t, S(t)), t \geq 0$[19]. Then $\hbar(\mathrm{t}, \mathrm{x}), t \geq 0, x > 0$, satisfies the PDE

---

[19] $\hbar(\mathrm{t}, \mathrm{x}), t \geq 0, x > 0$ has continuous $\frac{\partial \hbar(\mathrm{t},\mathrm{x}),}{\partial t}$ and $\frac{\partial^2 \hbar(\mathrm{t},\mathrm{x})}{\partial x^2}$, denoted by $\hbar \in C^{(1,2)}\big[[0,\infty) \times (0,\infty)\big]$.

(2.22) $$\frac{\partial \hbar(t,\mathrm{x})}{\partial t}+r\frac{\partial \hbar(t,\mathrm{x})}{\partial x}x-r\hbar(t,\mathrm{x})+\frac{1}{2}\frac{\partial^2 \hbar(t,\mathrm{x})}{\partial \mathrm{x}^2}x^2(\sigma^2+r\hat{c}^2)=0.$$

As expected, the introduction of cost in the hedging portfolio increases the volatility in the BSM-equation.

The next example shows that in a diffusion market with arbitrage opportunities the introduction of arbitrage taxes will eliminate any such opportunities. Consider a diffusion market with two risky assets:

$(i)$ $\mathcal{V}$ with price process: $V(t)=e^{\left(m-\frac{1}{2}v^2\right)t+\sigma W(t)}, t\geq 0, m\in R, \sigma>0$;

$(ii)$ $\mathcal{S}$ with price process: $S(t)=e^{(\mu-\frac{1}{2}\sigma^2)t+\sigma W(t)}, t\geq 0, \mu\neq m$.

Let $\mathcal{G}$ t be a perpetual derivative $\mathcal{G}$ with price process $g\big(S(t),V(t)\big), t\geq 0, g\in C^{(2,2)}[(0,\infty)^2]$. Then, standard replication arguments lead to the following PDFs for $g(x,y), x>0, y>0$,

(2.23) $$g(x,y)=\frac{\partial g(x,y)}{\partial x}x+\frac{\partial g(x,y)}{\partial y}y,$$

(2.24) $$\frac{\partial^2 g(x,y)}{\partial x^2}x^2+2\frac{\partial^2 g(x,y)}{\partial x\partial y}xy+\frac{1}{2}\frac{\partial^2 g(x,y)}{\partial y^2}y^2=0.$$

Consider $\mathcal{G}^{(a,b)}, a\in R, b\in R$, with price process $g^{(a,b)}\big(S(t),V(t)\big)=S(t)^aV(t)^b$. Then $g^{(a,b)}(x,y), x>0, y>0$, satisfies (2.23) and (2.24) if and only if $b=1-a$. Furthermore, $\mathcal{G}^{(a,1-a)}$ with price process $g^{(a,1-a)}\big(S(t),V(t)\big)=S(t)^aV(t)^{1-a}$, is a perpetual derivative generated by a self-financing strategy

$$a(t) = \frac{\partial g\big(S(t), V(t)\big)}{\partial x} S(t) = aS(t)^a V(t)^{1-a},$$

$$b(t) = \frac{\partial g\big(S(t), V(t)\big)}{\partial y} V(t) = bS(t)^a V(t)^{1-a}.$$

Consider then the following self-financing portfolio:

$$g^{(arb)}\big(S(t), V(t)\big) := g^{(1,0)}\big(S(t), V(t)\big) - 2g^{\left(\frac{1}{2},\frac{1}{2}\right)}\big(S(t), V(t)\big) + g^{(1,1)}\big(S(t), V(t)\big)$$

$$= S(t) - 2\sqrt{S(t)V(t)} + V(t) = \left(\sqrt{S(t)} - \sqrt{V(t)}\right)^2.$$

Thus, $g^{(arb)}\big(S(0), V(0)\big) = 0$, while $g^{(arb)}\big(S(0), V(0)\big) = \left(\sqrt{S(t)} - \sqrt{V(t)}\right)^2$ $\mathbb{P}$-a.s.

Consider now the same market $(\mathcal{V}, \mathcal{S})$, but assume that hedging can be performed with the class of Markov strategies with arbitrage taxes. Then, for the perpetual derivative $\mathcal{G}$ with price process $g\big(S(t), V(t)\big), t \geq 0$, let $\big(a(t), b(t)\big), t \geq 0$, be a Markov strategy with arbitrage taxes, that is, $g\big(S(t), V(t)\big) = a(t)S(t) + b(t)V(t)$ and

$$dg\big(S(t), V(t)\big) = \left\{a(t) - \frac{1}{2} c^{(S)^2} \frac{\partial^2 g\big(t, S(t), V(t)\big)}{\partial x^2} S(t)\right\} dS(t) +$$

$$+ \left\{ b(t) - \frac{1}{2} c^{(V)^2} \frac{\partial^2 g\big(t, S(t), V(t)\big)}{\partial y^2} V(t)\right\} dV(t),$$

for some tax-intensities $c^{(S)} > 0$ and $c^{(V)} > 0$. Then, $g(x, y), x \geq 0, y \geq 0$, satisfies the PDE

$$0 = \frac{\partial g(x, y)}{\partial x} x + \frac{\partial g(x, y)}{\partial y} y - g(x, y) + \frac{1}{2} c^{(S)^2} \frac{\partial^2 g(x, y)}{\partial x^2} x^2 + \frac{1}{2} c^{(V)^2} \frac{\partial^2 g(x, y)}{\partial y^2} y^2.$$

The boundary condition $g(1,1) = 0$ implies $g(x, y) = 0$. Thus, $(\mathcal{V}, \mathcal{S})$ is free of arbitrage opportunities in the class of Markov strategies with arbitrage taxes.

## 3. MIXED HERMITE MARKET MODEL WITH FRACTIONAL STOCHASTIC DRIFT

We now introduce the dynamics of the risky asset in a mixed Hermite market.[20] The dynamics of the risky asset should have the following five characteristics:

$(i)$ Various fractional market models suggested in the literature should be encompassed as special cases.[21]

$(ii)$ The model should lead to a market with no arbitrage opportunities (at least after the introduction of a strategy-specific arbitrage tax).[22]

---

[20] See for example, Cheridito (2001, 2003), Mishura (2008), and Kozachenko, Melnikov and Mishura (2014).

[21] See Mishura and Valkeida (2002), Mishura (2008), Bender, Sottinen and Valkeida (2011), Biagini et al. (2008), Kuznezov (1999), Cheridito (2001, 2003), Zähle (1998, 2002), and Rostek (2009).

[22] Similar to the proof of Propositions 1 and 2, see also Mishura (2008, Section 5.1.4), one can show that the mixed Hermit market with (i) riskless bond $\mathcal{M}$ with price processes (2.14), and (ii) risky asset $\mathcal{V}$ with price process $V(t) = \exp\{\mu t + \sigma W(t) + \sigma^{(H)} \mathcal{H}(t)\},\ t \geq 0,\ \mu \in R, \sigma > 0,\ \sigma^{(H)} > 0$, is free of arbitrage opportunities within the class of Markov strategies. In this paper, we are interested in the more general framework by considering the market $\left(\mathcal{M}, \mathbb{Z}^{(b)}, \mathbb{H}^{(\rho)}\right)$. There are two reasons for omitting Hermite markets with no arbitrage opportunities. The first is

$(iii)$ The return processes should have a flexible distributional tail behavior: from Gaussian (thin-tailed) to heavy tailed non-Gaussian distributions.

$(iv)$ The mean return process should grow linearly in time in order to be comparable to the dynamics of the riskless bond.

$(v)$ The model should be sufficiently flexible to provide close fits to real asset data while still being parsimonious.

We denote the risky asset in the Hermite market by $\mathbb{S}^{(H)}$ and assume that the $\mathbb{S}^{(H)}$-dynamics follow a mixed geometric Hermite process given by

(3.1) $S^{(H)}(t) = S^{(H)}(0) \exp\{\mu t + \sigma W(t) + \sigma^2(t^{1-2H}\mathcal{H}(t)^2 - t) + \sigma^{(H)}\mathcal{H}(t)\}, t \geq 0,$

$S^{(H)}(0) = x^{(H)}(0) > 0, \mu \in R, w > 0, \sigma^{(H)} > 0,$ and $W(t),\ t \geq 0,$ being a standard BM independent of the processes $\mathcal{H}(t), t \geq 0,$ defined by (2.2). This significantly generalizes the existing market models with LRD.[23]

Itô's formula and pathwise integration formula (A1.14), it follows that that the introduction of an arbitrage tax transforms both fractal and diffusion markets with arbitrage opportunities into arbitrage-free markets. The second is that regardless of how small the tax is, its introduction transforms Hermite markets with arbitrage opportunities into arbitrage-free markets.

[23] See, for example, Mishura (2008), Biagini et al. (2008), Rostek (2009), and Torres and Tudor (2009).

(3.2) $$\frac{dS^{(H)}(t)}{S^{(H)}(t)} = \left(\mu - \frac{1}{2}\sigma^2 + \sigma^2(1-2H)t^{-2H}\mathcal{H}(t)^2\right)dt + \sigma dW(t) +$$

$$+\left(\sigma^2 t^{1-2H} 2\mathcal{H}(t) + \sigma^{(H)}\right) d^{(S)}\mathcal{H}(t).$$

In (3.1) and (3.2),

$\boldsymbol{D}^{(\mathbf{1})}$: $\mu \in R$, is $\mathbb{S}^{(H)}$- instantaneous mean return;[24]

$\boldsymbol{D}^{(\mathbf{2})}$: $\sigma > 0$ is $\mathbb{S}^{(H)}$- diffusion volatility;[25]

$\boldsymbol{D}^{(\mathbf{3})}$: $\sigma^2(t^{1-2H}\mathcal{H}(t)^2 - t)$ $t \geq 0$, is $\mathbb{S}^{(H)}$- fractional stochastic drift;[26]

---

[24] This term is needed, as we would like to have a riskless asset (a bond), denoted by $\mathcal{M}$, with price dynamics $\beta(t) = \beta(0)e^{rt}t \geq 0, r \in R$.

[25] If $\sigma = 0$, a simple arbitrage strategy is as follows. Consider a "pure" Hermite market with zero interest rate: $S(t) = e^{\mathcal{H}^{(H,\mathscr{k})}(t)}, \beta(t) = 1, t \geq 0$. Form a strategy, $a(t) = 2S(t) - 2, b(t) = 1 - S(t)^2$, generating a self-financing portfolio $P^{(H,\mathscr{k})}(t) \coloneqq a(t)S(t) + b(t)\beta(t) = (S(t)-1)^2$. Clearly, $P^{(H,\mathscr{k})}(0) = 0$, while $P^{(H,\mathscr{k})}(t) > 0, \mathbb{P}$-a.s. Shiryayev (1998) provided this example for the case of a FBM, $B^H = \mathcal{H}^{(H,1)}$. Note that $a(t), b(t), t \geq 0$, are smooth functions of the stock price $S(t)$ only, that is, they are Markov self-financing strategies.

[26] Because, $(i)\mathbb{E}\mathcal{H}(t) = \sum_{i=1}^{I}\alpha^{(i)}\mathbb{E}\mathcal{H}^{(H,\mathscr{k}(i))}(t) = 0, \mathbb{E}\mathcal{H}(t)^2 = \sum_{i=1}^{I}{\alpha^{(i)}}^2\mathcal{H}^{(H,\mathscr{k}(i))}(t) = t^{2H}, H \in \left(\frac{1}{2}, 1\right)$, and $(ii)$ $\mathcal{H}^{(H,\mathscr{k})}(t)^2, \mathscr{k} \in \mathbb{N}$, has heavy-tailed marginal distributions when $\mathscr{k}$ increases, see $\boldsymbol{\mathcal{HBP}}^{(\mathbf{9})}$ (in Appendix A1), then the term $\sigma^2(t^{1-2H}\mathcal{H}(t)^2 - t), t \geq 0$, adds flexibility to the price dynamics when fitted in the real data.

$\boldsymbol{D}^{(4)}$: $\sigma^{(H)} > 0$ is $\mathbb{S}^{(H)}$- fractional volatility.

As riskless asset we choose $\mathcal{M}$ with price dynamics $\beta(t) = \beta(0)e^{rt}, t \geq 0, r \in R, \beta(0) > 0$. Let $\mathbb{Z}^{(b)}, b > 0$, be a risky asset with price dynamics $Z(t), t \geq 0$ following a GBM with zero drift

(3.3) $$Z^{(b)}(t) = e^{-\frac{1}{2}b^2 t + bW(t)}, t \geq 0.$$

Let $\mathbb{H}^{(\rho)}, \rho > 0$, be a risky asset with mixed geometric Hermite process

(3.4) $$Y^{(\rho)}(t) = \exp\{W(t) + (t^{1-2H}\mathcal{H}(t)^2 - t) + \rho\mathcal{H}(t)\}, t \geq 0.$$

We refer to $\mathcal{M}, \mathbb{Z}^{(b)}$, and $\mathbb{H}^{(\rho)}$ as **basic assets** in mixed Hermite markets.[27] We shall show that the market $\left(\mathcal{M}, \mathbb{Z}^{(b)}, \mathbb{H}^{(\rho)}\right)$ admits arbitrage opportunities but is free of such opportunities when an arbitrage tax is included.

We shall consider only self-financing strategies (**SFSs**) $(b_t, z_t, y_t,), t \geq 0$, which are **SFSs of Markov-type**, that is, generating **self-financing portfolios of Markov-type**

(3.5) $$P^{(M)}(t) = P^{(M)}(0) + \int_0^t \mathfrak{b}_s d\beta(s) + \int_0^t \mathfrak{z}_s dZ^{(b)}(s) + \int_0^t \mathfrak{y}_s d^{(S)}Y^{(\rho)}(s), t \geq 0,$$

where

[27] In all studies of fractal markets, the goal is to extend a given diffusion market model which is arbitrage-free to a fractal market which under certain conditions can be arbitrage-free. The market $\left(\mathcal{M}, \mathbb{Z}^{(b)}, \mathbb{H}^{(\rho)}\right)$ is an exception: if we set in (3.4) $H = \frac{1}{2}, b \neq 2$, and replace $\mathcal{H}(t), t \geq 0$, with a BM $B(t), t \geq 0$ independent of $W(t), t \geq 0$, the market $\left(\mathcal{M}, \mathbb{Z}^{(b)}, \mathbb{H}^{(\rho)}\right)$ will have arbitrage opportunities.

$(i)$ $\quad P^{(M)}(t) = \mathcal{P}^{(M)}\left(t, Z^{(b)}(t), Y^{(\rho)}(t)\right), \mathfrak{b}_t = \mathfrak{b}\left(t, Z^{(b)}(t), Y^{(\rho)}(t)\right);$

$(ii)$ $\quad \mathfrak{z}_t = \mathfrak{z}\left(t, Z^{(b)}(t), Y^{(\rho)}(t)\right),\ \mathfrak{y}_t = \mathfrak{h}\left(t, Z^{(b)}(t), Y^{(\rho)}(t)\right), \mathcal{P}^{(M)}(t, z, y);$

$(iii)$ $\quad P^{(M)}(t, x, y, ), \mathfrak{b}(t, z, y), \mathfrak{z}(t, x, y), \mathfrak{h}(t, z, y), t \geq 0, x > 0, y > 0$

have continuous derivatives of first order with respect to $t \geq 0$, and of second order with respect to $x > 0, y > 0$.

*PROPOSITION 3. Within the class of Markov self-financing strategies, the corresponding portfolios satisfying (3.5) will also satisfy the PDE*

*(3.6)* $$\frac{\partial \mathcal{P}^{(M)}(t,x,y)}{\partial t} + rx\frac{\partial \mathcal{P}^{(M)}(t,x,y)}{\partial x} + ry\frac{\partial \mathcal{P}^{(M)}(t,x,y)}{\partial y} - r\mathcal{P}^{(M)}(t, x, y) = 0.$$

*As a consequence, the market* $\left(\mathcal{M}, \mathbb{Z}^{(b)}, \mathbb{H}^{(\rho)}\right)$ *admits arbitrage opportunities.*

*Proof:* Similar to that of Proposition 1 and omitted.

Next, let us suppose that the trading Markov strategy in a Hermite market comes at a cost. Therefore, we require the following portfolio dynamics:

(3.7) $$P^{(M)}(t) = \mathcal{P}^{(M)}\left(t, Z^{(b)}(t), Y^{(\rho)}(t)\right) = \mathfrak{b}_t\beta(t) + \mathfrak{z}_t Z^{(b)}(t) + \mathfrak{y}_t Y^{(\rho)}(t)$$

with embedded running cost

(3.8) $$P^{(M)}(t) - P^{(M)}(0) = \int_0^t \mathfrak{b}_s d\beta(s) +$$

$$+\int_0^t \left(\mathfrak{z}_t - \frac{1}{2}{c^{(Z)}}^2 \frac{\partial^2 \mathcal{P}^{(M)}\left(t, Z^{(b)}(t), Y^{(\rho)}(t)\right)}{\partial x^2} Z^{(b)}(t)\right) d^{(S)}Z^{(b)}(t) +$$

$$+\int_0^t \left( \mathfrak{y}_t - \frac{1}{2} c^{(Y)^2} \frac{\partial^2 \mathcal{P}^{(M)}\left(t, Z^{(b)}(t), Y^{(\rho)}(t)\right)}{\partial y^2} Y^{(\rho)}(t) \right) d^{(S)} Y^{(\rho)}(t)$$

for some $c^{(Z)} > 0,\ c^{(Y)} > 0.$

*PROPOSITION 4.* $\left(\mathcal{M}, \mathbb{Z}^{(b)}, \mathbb{H}^{(\rho)}\right)$ *is free of arbitrage within the class of Markov strategies with cost given by (3.7) and (3.8)*

*Proof:* Similar to that of Proposition 2 and omitted.

## 5. CONCLUSIONS

In this paper, we argue that the following two principles should be followed when studying fractal markets, such as the general class of Hermite markets. First, in Hermite markets, the riskless asset dynamics should not be imposed, but derived as the price process of a perpetual derivative of the risky assets constituting the Hermite market. Second, Hermite markets are inherently clairvoyant and therefore an arbitrage tax can be imposed on the trading strategies. We introduce a specially designed arbitrage tax, proportional to the velocity of the hedging portfolio, such that regardless of how small it is, the Hermite market becomes free of arbitrage opportunities. We also show that, even in the classical case of diffusion markets with arbitrage opportunities, the introduction of this new tax can make such diffusion markets arbitrage free.

## APPENDIX 1: HERMITE MOTION AS A MODEL FOR MARKET UNCERTAINTY

**Hermite motion** (HM),**[28]** $\mathcal{H}^{(H,\mathscr{k})}(t), t \geq 0, \mathscr{k} \in \mathbb{N}, H \in (\frac{1}{2}, 1)$ is defined by

(A1.1) $\mathcal{H}^{(H,\mathscr{k})}(t) = C^{(H,\mathscr{k})} \int_{\mathcal{D}^{\mathscr{k}}} K_t^{(H,\mathscr{k})}\left(v^{(1)}, \ldots, v^{(\mathscr{k})}\right) dB\left(v^{(1)}\right) \ldots dB\left(v^{(\mathscr{k})}\right), t \geq 0,$[29]

where

$(\boldsymbol{HMi})$ $\mathcal{D}^{\mathscr{k}} \coloneqq \left\{\mathbb{v} = \left(v^{(1)}, \ldots, v^{(\mathscr{k})}\right) \in R^{\mathscr{k}} : v^{(i)} \neq v^{(j)}, i, j = 1, \ldots, \mathscr{k}, i \neq j\right\}$;

$(\boldsymbol{HMii})$ For a given $t \geq 0,$ the kernel $K_t^{(H,\mathscr{k})}(\mathbb{v}), t \geq 0, \mathbb{v} = \left(v^{(1)}, \ldots, v^{(\mathscr{k})}\right) \in R^{\mathscr{k}}$, is defined by[30]

(A1.2) $$K_t^{(H,\mathscr{k})}(\mathbb{v}) \coloneqq \int_0^t \left[\prod_{j=1}^{\mathscr{k}} \left(s - v^{(j)}\right)_+^{\frac{H-1}{\mathscr{k}} - \frac{1}{2}}\right] ds$$

---

[28] See Taqqu (1979), Dobrushin (1979), Dobrushin and Major (1979), Dehling and Taqqu (1989), Lacey (1991), Embrechts and Maejima (2002), Lavancier (2006), Maejima and Tudor (2007), Chronopoulou (2008), Tudor (2008), Torres and Tudor (2009), Pipiras and Taqqu (2010), Chronopoulou, Tudor and Viens (2011), Tudor (2013), Marty (2013), Bai and Taqqu (2014), Sun and Cheng (2014), Clausel et al. (2014), and Fauth and Tudor (2016).

[29] The integral is understood as a multiple Wiener-Itô integral, see, for example, Dobrushin (1979), Nualart (2006), and Clausel et al. (2014).

[30] $a_+^b \coloneqq \begin{cases} a^b, if\ a \geq 0 \\ 0, if\ a < 0 \end{cases}, b \in R$. For every given $t \geq 0$, the kernel $K_t^{(H,\mathscr{k})}(\mathbb{v})$, is symmetric and has a finite $\mathcal{L}_2\left(R^{\mathscr{k}}\right)$-norm: $\left\|K_t^{(H,\mathscr{k})}\right\|_{\mathcal{L}_2\left(R^{\mathscr{k}}\right)} = \sqrt{\int_{R^{\mathscr{k}}} \left(K_t^{(H,\mathscr{k})}(\mathbb{v})\right)^2 d\mathbb{v}} < \infty$ . Thus, $\mathcal{H}^{(H,\mathscr{k})}(t), t \geq 0$ is well-defined process.

$(\boldsymbol{HMiii})$ $H \in \left(\frac{1}{2}, 1\right)$ is the Hurst index (index of self-similarity);[31]

$(\boldsymbol{HMiv})$ $C^{(H,k)} > 0$ is a normalizing constant such that $\mathbb{E}\left(\mathcal{H}^{(H,k)}(1)\right)^2 = 1$.[32]

$(\boldsymbol{HMv})$ $B(v), v \in R$ is a two-sided BM[33] defined on $(\Omega, \mathcal{F}, \{\mathcal{F}_t\}_{t\geq 0}, \mathbb{P})$.

An alternative representation,[34] $\mathcal{H}^{(H,k)}(t), t \geq 0$, is given by

(A1.3)
$$\mathcal{H}^{(H,k)}(t) = c^{(H,k)} \int_{R^k} \frac{e^{it\sum_{j=1}^{k} u^{(j)}} - 1}{i\left[\sum_{j=1}^{k} u^{(j)}\right]\left|\prod_{j=1}^{k} u^{(j)}\right|^{\frac{2H-2+k}{2k}}} B^{(\mathbb{C})}\left(du^{(1)}\right) \dots B^{(\mathbb{C})}\left(du^{(k)}\right),$$

where $c^{(H,k)}$ is a normalizing constant, so that $\mathbb{E}\left(\mathcal{H}^{(H,k)}(1)\right)^2 = 1$, and $B^{(\mathbb{C})}(du)$ is a complex random measure generated by a standard Brownian motion.

---

[31] See, Samorodnitsky (2016) for an extensive study of LRD processes.

[32] $C^{(H,k)} = \left(\sqrt{k!}\left\|K_1^{(H,k)}(\mathbb{v})\right\|_{\mathcal{L}_2(R^k)}\right)^{-1}$, $C^{(H,1)} = \sqrt{\frac{2H\Gamma\left(\frac{3}{2}-H\right)}{\Gamma\left(\frac{1}{2}+H\right)\Gamma(2-2H)}}$ and $C^{(H,2)} = \frac{\Gamma\left(1+\frac{H}{2}\right)\sqrt{\frac{H}{2}(2H-1)}}{\Gamma(\frac{H}{2})\Gamma(1-H)}$, see Clausel et al. (2014).

[33] The two-sided Brownian motion $B(v), v \in R$ is defined as follows:

$$B(v) = \begin{cases} B^{(1)}(v), for\ v \geq 0 \\ B^{(2)}(-v), for\ v < 0, \end{cases}$$

where $B^{(1)}(t), t \geq 0$ and $B^{(2)}(t), t \geq 0$ are two independent standard BMs.

[34] See Taqqu (1979).

The existence of the HM follows from the non-central invariance principle[35]

(A1.4) $$\frac{1}{H}\sum_{j=1}^{\lfloor nt\rfloor} \mathbb{g}^{(\hbar)}\left(\xi^{(j)}\right) \to^{(weakly)} \left(\mathcal{H}^{(H,\hbar)}(t)\right)_{0\le t\le T},$$

where $(i)$ $\xi^{(j)}, j = 0, \pm 1, \pm 2 \ldots$, is a stationary Gaussian sequence, with $\mathbb{E}\xi^{(j)} = 0$, $\mathbb{E}\left(\xi^{(j)}\right)^2 = 1$ and covariance function $\rho^{(\xi)}(n) = \mathbb{E}\left[\xi^{(0)}\xi^{(n)}\right]$ having power decay for some slowly varying function[36] $L(n), n = 1,2, \ldots$,

(A1.5) $$\lim_{n\uparrow\infty} \frac{\rho^{(\xi)}(n)}{L(n)n^{\frac{2H-2}{\hbar}}} < \infty;$$

$(ii)$ $\mathbb{g}^{(\hbar)}: R \to R$, where $\mathbb{E}\mathbb{g}^{(\hbar)}\left(\xi^{(0)}\right) = 0, \mathbb{E}\left(\mathbb{g}^{(\hbar)}\left(\xi^{(0)}\right)\right)^2 < \infty$, and $\mathbb{g}^{(\hbar)}$ has Hermite rank $\hbar$.[37]

The basic properties of the HM, $\mathcal{H}^{(H,\hbar)} = \left\{\mathcal{H}^{(H,\hbar)}(t), t \ge 0\right\}$, are:

---

[35] See Dobrushin and Major (1979), Taqqu (1979), and Torres and Tudor (2009).

[36] See Seneta (1976).

[37] Let $g^{(m)}(x) = (-1)^m e^{\frac{x^2}{2}} \frac{d}{dx} e^{-\frac{x^2}{2}}, x \in R,$ be a Hermite polynomial of order $m = 0,1, \ldots$ The function $\mathbb{g}^{(\hbar)}: R \to R$ has Hermite rank $\hbar$, if $\mathbb{g}^{(\hbar)}$ has the following representation:

$$\mathbb{g}^{(\hbar)}(x) = \sum_{m\ge 0} c^{(m)}\, g^{(m)}(x), c^{(m)} := \frac{1}{m!}\mathbb{E}\left\{\mathbb{g}^{(\hbar)}\left(\xi^{(0)} g^{(m)}\left(\xi^{(0)}\right)\right)\right\},$$

with $\hbar = \min\left\{m: c^{(m)} \ne 0\right\}$, see Torres and Tudor (2009).

$\boldsymbol{\mathcal{HBP}^{(1)}}$: $\mathcal{H}^{(H,1)}(t) = B^H(t), t \geq 0$, is a FBM and, thus, a Gaussian process. For every $\mathscr{k} \geq 2$, $\mathcal{H}^{(H,\mathscr{k})}$[38] is a non-Gaussian process. For all $\mathscr{k} \in \mathbb{N}$, $\mathcal{H}^{(H,\mathscr{k})}$ is neither a semimartingale nor a Markov process.

$\boldsymbol{\mathcal{HBP}^{(2)}}$: For all $\mathscr{k} \in \mathbb{N}, H \in \left(\frac{1}{2}, 1\right), \mathcal{H}^{(H,\mathscr{k})}$ trajectories are nowhere differentiable. However, for every $0 < \alpha < H$, $\mathcal{H}^{(H,\mathscr{k})}$ trajectories are $\alpha$-Hölder continuous ($\alpha$-Lipchitz continuous), that is, for some $C(H,\mathscr{k}) > 0$:

(A1.6) $\left|\mathcal{H}^{(H,\mathscr{k})}(t) - \mathcal{H}^{(H,\mathscr{k})}(s)\right| \leq C(H,\mathscr{k})|t-s|^{\alpha}$, for all $t \geq 0, s \geq 0$.

$\boldsymbol{\mathcal{HBP}^{(3)}}$: The trajectories of $\mathcal{H}^{(H,\mathscr{k})}$ are of bounded *$\Phi$- variation*,[39] where

$$\Phi(\mathrm{u}) = \frac{\mathrm{u}^{1/\mathrm{H}}}{\left[\log\left(\log\left(\frac{1}{\mathrm{u}}\right)\right)\right]^{\frac{\mathscr{k}}{2H}}}, u > 0.$$ [40]

$\boldsymbol{\mathcal{HBP}^{(4)}}$: Let $V^{(N)}(\mathcal{H}^{(H,\mathscr{k})})$ be the *centered quadratic variation* of $\mathcal{H}^{(H,\mathscr{k})}$, i.e.,

---

38 $\mathcal{H}^{(H,2)}$ is known as the Rosenblatt process, see Taqqu (2011).

39 The $\Phi$- variation of a function $f: [0,1] \to R$ is defined by

$V^{(\Phi)}(f) \coloneqq limsup_{\{\mathcal{P}^{(n)}, d(\mathcal{P}^{(n)}) \to 0, n\uparrow\infty\}} \sum_{k=1,\dots,n,(t^{(0)},t^{(1)},\dots,t^{(n)})\in\mathcal{P}^{(n)}} \Phi\left(\left|f\left(t^{(k)}\right) - f(t^{(k-1)})\right|\right)$,

where $\mathcal{P}^{(n)} \coloneqq \left\{\left(t^{(0)}, t^{(1)}, \dots, t^{(n)}\right): 0 = t^{(0)} < t^{(1)} < \dots < t^{(n)} = 1\right\}$ *and* $d\left(\mathcal{P}^{(n)}\right) \coloneqq max_{k=1,\dots,n}\left|t^{(k)} - t^{(k-1)}\right|$.

40 See Basse-O'Connor and Weber (2016).

(A1.7) $$V^{(N)}\left(\mathcal{H}^{(H,k)}\right)=\sum_{n=0}^{N}\left\{\begin{array}{c}\left(\mathcal{H}^{(H,k)}\left(t^{(n+1)}\right)-\mathcal{H}^{(H,k)}\left(t^{(n)}\right)\right)^{2}- \\ -\mathbb{E}\left[\left(\mathcal{H}^{(H,k)}\left(t^{(n+1)}\right)-\mathcal{H}^{(H,k)}\left(t^{(n)}\right)\right)^{2}\right]\end{array}\right\},$$

where $t^{(n)}-t^{(n-1)}=\gamma^{(N)}>0, t^{(0)}=0$; and let

(A1.8) $$\delta^{(N)}\left(\mathcal{H}^{(H,k)}\right):=\sqrt{\mathbb{E}\left[\left(V^{(N)}(\mathcal{H}^{(H,k)})\right)^{2}\right]}.$$

Then,[41]

(**I**) if $k=1$ and $\frac{1}{2}<H\leq\frac{3}{4}$, then the limit $lim_{N\uparrow\infty}\ \frac{\delta^{(N)}\left(\mathcal{H}^{(H,k)}\right)}{\left(\gamma^{(N)}\right)^{2H}\sqrt{N}}<\infty$ exists and is finite, and

(A1.9) $$\frac{V^{(N)}\left(\mathcal{H}^{(H,k)}\right)}{\delta^{(N)}\left(\mathcal{H}^{(H,k)}\right)}\Rightarrow^{distr}\boldsymbol{\mathcal{N}}(0,1);$$

(**II**) if $k=1$ and $\frac{3}{4}<H<1$, then the limit $lim_{N\uparrow\infty}\ \frac{\delta^{(N)}\left(\mathcal{H}^{(H,k)}\right)}{\left(\gamma^{(N)}\right)^{2H}N^{2H-1}logN}<\infty$ exists and is finite, and

(A1.10) $$\frac{V^{(N)}\left(\mathcal{H}^{(H,k)}\right)}{\delta^{(N)}\left(\mathcal{H}^{(H,k)}\right)}\Rightarrow^{distr}\left(\boldsymbol{\mathcal{R}}^{(H)}(0,1)\right)^{1-\frac{2(1-H)}{k}};$$

(**II**) if $k>1, H\in\left(\frac{1}{2},1\right)$, then the limit $lim_{N\uparrow\infty}\ \frac{\delta^{(N)}\left(\mathcal{H}^{(H,k)}\right)}{\left(\gamma^{(N)}\right)^{2H}N^{\left(1-\frac{2(1-H)}{k}\right)}}<\infty$ exists and is finite, and (A1.10) holds.

---

[41] See Clausel et al. (2016). In what follows: (i) $\Rightarrow^{distr}\boldsymbol{\mathcal{N}}(0,1)$ is a standard normal random variable; and (ii) $\mathcal{R}(0,1)$ is a standard Rosenblatt variable, that is, $\boldsymbol{\mathcal{R}}^{(H)}(0,1)$ has the same distribution as $\mathcal{H}^{(H,2)}(1)$.

$\boldsymbol{\mathcal{HBP}^{(5)}}$: $\mathcal{H}^{(H,\hbar)}$ has stationary increments, zero mean, variance $\mathbb{E}\left[\left(\mathcal{H}^{(H,\hbar)}(t)\right)^2\right] = t^{2H}$, and the covariance function $\rho^{(H,\hbar)}$ is given by

(A1.11) $\rho^{(H,\hbar)}(t,s) = \mathbb{E}\left(\mathcal{H}^{(H,\hbar)}(t)\mathcal{H}^{(H,\hbar)}(s)\right) = \frac{1}{2}(t^{2H} + s^{2H} - |t-s|^{2H}\}), s \geq 0, t \geq 0;$

$\boldsymbol{\mathcal{HBP}^{(6)}(Long-range\ dependence\ (LRD))}$: Let $\Delta^{(H,\hbar)}(n) \coloneqq \mathcal{H}^{(H,\hbar)}(n+1) - \mathcal{H}^{(H,\hbar)}(n)$, $n = 0,1,\ldots,$ be the sequence of unit increments of $\mathcal{H}^{(H,\hbar)}$. Then, $lim_{n\uparrow\infty} n^{2-2H}\mathbb{E}\left(\Delta^{(H,\hbar)}(n)\Delta^{(H,\hbar)}(0)\right) = H(2H-1)$, and, in particular,

(A1.12) $$\sum_{n=1}^{\infty} \mathbb{E}\left(\Delta^{(H,\hbar)}(n)\Delta^{(H,\hbar)}(0)\right) = \infty.$$

$\boldsymbol{\mathcal{HBP}^{(7)}}$: $\mathcal{H}^{(H,\hbar)}$ is a self-similar process with Hurst index $H$,

(A1.13) $$\mathcal{H}^{(H,\hbar)}(ct) \triangleq c^H \mathcal{H}^{(H,\hbar)}(t), t \geq 0, c > 0.$$

The smoothness of $\mathcal{H}^{(H,\hbar)\prime}s$ trajectories allows us to define stochastic integrals with respect to $\mathcal{H}^{(H,\hbar)}$ in a pathwise sense.[42] Stochastic calculus with HM is based on the *fractional Stratonovich integral*: For a continuous function, $f:[0,T] \to R$, the Stratonovich integral is denoted by $\int_0^T f(s)\, d^{(S)}\mathcal{H}^{(H,\hbar)}(t)$ and defined as a limit of the Riemann sums[43]

[42] See, for example, Russo and Vallois (1993) and Neuenkirch and Nourdin (2007).

[43] This is because $H \in \left(\frac{1}{2}, 1\right)$, the integral $\int_0^T f(s) d^{(S)}\, \mathcal{H}^{(H,\hbar)}(s) =$

$= lim_{n\uparrow\infty} \sum_{k=0, t^{(k)}=\frac{k}{n}T, k=0,\ldots,n}^{n-1} f\left((1-\delta)t^{(k)} + \delta t^{(k+1)}\right)\left(\mathcal{H}^{(H,\hbar)}(t^{(k+1)}) - \mathcal{H}^{(H,\hbar)}(t^{(k)})\right)$

has the same value for all $\delta \in [0,1]$; see Duncan (2000).

(A1.14) $\int_0^T f(s)\, d^{(S)}\mathcal{H}^{(H,\&)} =$

$$= lim_{n\uparrow\infty\,\left(t^{(k)}=\frac{k}{n}T,k=0,\dots,n\right)} \sum_{k=0}^{n-1} f\left(t^{(k)}\right)\left(\mathcal{H}^{(H,\&)}(t^{(k+1)}) - \mathcal{H}^{(H,\&)}(t^{(k)})\right).$$

This implies the following chain-rule: given a sufficiently smooth function $G(x,t), x \in R, t \geq 0$,

(A1.15) $$G\left(\mathcal{H}^{(H,\&)}(t+s), t+s\right) = G\left(\mathcal{H}^{(H,\&)}(t), t\right) +$$

$$\int_t^{t+s} \frac{\partial G\left(\mathcal{H}^{(H,\&)}(u), u\right)}{\partial x} d^{(S)}\mathcal{H}^{(H,\&)}(t) + \int_t^{t+s} \frac{\partial G\left(\mathcal{H}^{(H,\&)}(u), u\right)}{\partial u} du,$$

or, in differential terms,[44]

$$dG\left(\mathcal{H}^{(H,\&)}(t), t\right) = \frac{\partial G(\mathcal{H}^{(H,\&)}(t),t)}{\partial x} d^{(S)}\mathcal{H}^{(H,\&)}(t) + \frac{\partial G\left(\mathcal{H}^{(H,\&)}(t),t\right)}{\partial t} dt.$$

$\boldsymbol{\mathcal{H}BP^{(8)}}$ (***Finite time interval representation***): The next representation is heuristic, as the trajectories of the Fractional Brownian Motion (FBM), $B^{(H)}(t) = \mathcal{H}^{(H,1)}(t), t \geq 0$, are not differentiable, but can be made precise using generalized functions:

(A1.16) $$\mathcal{H}^{(H,\&)}(t) = C^{(H,\&)} \int_{\mathcal{D}^{\&}} \int_0^t \left[\prod_{j=1}^{\&} \left(s - v^{(j)}\right)_+^{H^{(\&)}-\frac{3}{2}}\right] ds\, dB\left(v^{(1)}\right) \dots dB\left(v^{(\&)}\right)$$

$$= C^{(H,\&)} \int_0^t \dot{B}^{(H^{(\&)})}(s)^{\&}\, ds,$$

---

[44] In the literature, two alternative notations are used instead of $\dots d^{(S)}\mathcal{H}^{(H,\&)}(t)$:

$(i)$ $\dots.\delta d\mathcal{H}^{(H,\&)}(t)$ ; $(ii)$ $(S)$ $\dots. d\mathcal{H}^{(H,\&)}(t)$; $(iii) \dots d^{-}\mathcal{H}^{(H,\&)}(t)$.

where $H^{(\mathscr{k})} = \frac{H-1}{\mathscr{k}} + 1$[45]. The derivative $\dot{B}^{(H^{(\mathscr{k})})}(t) = \frac{\partial}{\partial s} B^{(H^{(\mathscr{k})})}(t), t \geq 0$, is known as **fractional white noise**. Thus, we can define the Hermite white noise $\dot{\mathcal{H}}^{(H,\mathscr{k})}(t), \mathcal{H}^{(H,\mathscr{k})}(t) = \int_0^t \dot{\mathcal{H}}^{(H,\mathscr{k})}(s)\, ds$, as follows:

(A1.17) $$\dot{\mathcal{H}}^{(H,\mathscr{k})}(t) = \dot{B}^{(H^{(\mathscr{k})})}(t)^{\mathscr{k}}, t \geq 0.$$

$\boldsymbol{\mathcal{H}BP^{(9)}}$(***Distributional tails of Hermite Process unit increment***):[46]

($\boldsymbol{i}$) There exist positive constants $a(H,\mathscr{k})$ and $b(H,\mathscr{k})$ such that for every $u > 0$,

$$\exp\left(-a(H,\mathscr{k})\, u^{\frac{2}{k}}\right) \leq \mathbb{P}\left(\left|\mathcal{H}^{(H,\mathscr{k})}(1)\right| \geq u\right) \leq \exp\left(-b(H,\mathscr{k})\, u^{\frac{2}{k}}\right)$$ [47];

($\boldsymbol{ii}$) There exist positive constants $c(H,\mathscr{k})$ and $d(H,\mathscr{k})$ such that for every $n \in \mathbb{N}$,

$$c(H,\mathscr{k})^n n\Gamma(\frac{kn}{2}); \leq \mathbb{E}\left(\left|\mathcal{H}^{(H,\mathscr{k})}(1)\right|^n\right) \leq d(H,\mathscr{k})^n n\Gamma(\frac{kn}{2})$$ [48];

$\boldsymbol{\mathcal{H}BP^{(10)}}$ (***Persistence of Hermite Processes***):[49] For all $\mathscr{k} \in \mathbb{N}$ and sufficiently large $T > 0$, there exists constant $\mathbb{c}^{(H,\mathscr{k})} > 0$,

---

[45] See Molchan (1969), Molchan and Golosov (1969), Taqqu (2011), and Tsoi (2011) for the two cases when $\mathscr{k} = 1$ and2.

[46] See Chen, Xu and Zhu (2015).

[47] Note that when $\mathscr{k}$ increases, the tails of distribution of the unit increment becomes heavier.

[48] By the Stirling's formula $\lim_{u\uparrow\infty} \frac{\Gamma(u+1)}{\sqrt{2\pi u}\left(\frac{u}{e}\right)^u} = 1$.

[49] See Aurzada and Mönch (2016).

$$\mathbb{P}\left(sup_{t\in[0,T]}\mathcal{H}^{(H,\mathcal{k})}(t)\leq 1\right)\geq \frac{1}{T^{1-H}(logT)^{\mathbb{c}(H,\mathcal{k})}}.$$

For $\mathcal{k}=1,2$ a sharper bound holds:

$$\mathbb{P}\left(sup_{t\in[0,T]}\mathcal{H}^{(H,\mathcal{k})}(t)\leq 1\right)\sim T^{H-1+o(1)}, \text{ as } T\uparrow\infty.$$

$\boldsymbol{\mathcal{H}BP^{(11)}}$ $(\boldsymbol{Hermite\ Ornstein-Uhlenbeck\ Process})$[50] The process $\mathcal{OU}^{(H,\mathcal{k})}(t), t\geq 0$,

$$\mathcal{OU}^{(H,\mathcal{k})}(t)=\sigma\int_{-\infty}^{t}e^{-\lambda(t-u)}d^{(S)}\mathcal{H}^{(H,\mathcal{k})}(u)\ ,t\geq 0,$$

with initial condition $\mathcal{OU}^{(H,\mathcal{k})}(0)=\ \sigma\int_{-\infty}^{0}e^{-\lambda(t-u)}d^{(S)}\,\mathcal{H}^{(H,\mathcal{k})}(u)$ is called **Hermite Ornstein-Uhlenbeck (HOU) process.** For all $\mathcal{k}\in\mathbb{N}$ , $\mathcal{OU}^{(H,\mathcal{k})}$ is a centered Gaussian process with covariance function

$$\mathbb{E}\mathcal{OU}^{(H,\mathcal{k})}(t)\mathcal{OU}^{(H,\mathcal{k})}(s)=\sigma^2\int_{-\infty}^{t}\int_{-\infty}^{s}e^{-\lambda(t-u)}e^{-\lambda(s-v)}|u-v|^{2H-2}dv\,du.$$

The HOU exhibits LRD. More precisely, as $s\uparrow\infty$,

$$\mathbb{E}\mathcal{OU}^{(H,\mathcal{k})}(t)\mathcal{OU}^{(H,\mathcal{k})}(t+s)\sim\frac{1}{2}\left(\frac{\sigma}{\lambda}\right)^2 2H(2H-1)\left\{s^{2H-2}-e^{-\lambda t}(t+s)^{2H-2}\right\}+O(s^{2H-4}).$$

## APPENDIX 2. PROOFS OF PROPOSITIONS

Appendix 2.1: Proof of Proposition 1

Consider a Markov self-financing strategy $\mathcal{P}\left(\mathcal{S}^{(PH)}(t)\right)=\sum_{i=1}^{N}a^{(i)}\left(\mathcal{S}^{(PH)}(t)\right)S^{(PH,i)}(t)\,,t\geq 0,$ and

[50] See Maejima and Tudor (2007).

$$\mathcal{P}\left(\mathcal{S}^{(PH)}(t)\right) - \mathcal{P}\left(\mathcal{S}^{(PH)}(0)\right) = \sum_{j=1}^{N}\int_0^t d^{(S)} a^{(i)}\left(\mathcal{S}^{(PH)}(s)\right) S^{(PH,i)}(s) =$$

$$= \sum_{j=1}^{N}\int_0^t a^{(i)}\left(\mathcal{S}^{(PH)}(s)\right) d^{(S)} S^{(PH,i)}(s).$$

Then, for all $t \geq 0$,

$$0 = \sum_{i=1}^{N}\int_0^t S^{(PH,i)}(s) d^{(S)} a^{(i)}\left(\mathcal{S}^{(PH)}(s)\right) =$$

$$\sum_{i=1}^{N}\int_0^t S^{(PH,i)}(s) \sum_{j=1}^{N} \frac{\partial a^{(i)}\left(\mathcal{S}^{(PH)}(s)\right)}{\partial x^{(j)}} S^{(PH,j)}(s) \left(\frac{\partial \mu^{(j)}(s)}{\partial s} ds + \mathcal{H}(s) \frac{\partial \sigma^{(j)}(s)}{\partial s} ds +\right.$$

$$\left.\sigma^{(j)}(s) d^{(S)} \mathcal{H}(s)\right)$$

Thus, for all $t \geq 0$,

$$\sum_{i=1}^{N}\left(\frac{\partial \mathcal{P}\left(\mathcal{S}^{(PH)}(t)\right)}{\partial x^{(j)}} - a^{(j)}\left(\mathcal{S}^{(PH)}(t)\right)\right) S^{(PH,j)}(t) \frac{\partial \mu^{(j)}(t)}{\partial t} = 0,$$

$$\sum_{i=1}^{N}\left(\frac{\partial \mathcal{P}\left(\mathcal{S}^{(PH)}(t)\right)}{\partial x^{(j)}} - a^{(j)}\left(\mathcal{S}^{(PH)}(t)\right)\right) S^{(PH,j)}(t) \frac{\partial \sigma^{(j)}(t)}{\partial t} = 0,$$

and

$$\sum_{i=1}^{N}\left(\frac{\partial \mathcal{P}\left(\mathcal{S}^{(PH)}(t)\right)}{\partial x^{(j)}} - a^{(j)}\left(\mathcal{S}^{(PH)}(t)\right)\right) S^{(PH,j)}(t) \sigma^{(j)}(t) = 0.$$

Thus, a solution of all those equations is $a^{(j)}\left(\mathcal{S}^{(PH)}(t)\right) = \frac{\partial \mathcal{P}\left(\mathcal{S}^{(PH)}(t)\right)}{\partial x^{(j)}}$. From, $\mathcal{P}\left(\mathcal{S}^{(PH)}(t)\right) = \sum_{i=1}^{N} a^{(i)}\left(\mathcal{S}^{(PH)}(t)\right) S^{(PH,i)}(t)$, it follows that $\mathcal{P}\left(\mathcal{S}^{(PH)}(t)\right) = \sum_{i=1}^{N} \frac{\partial \mathcal{P}\left(\mathcal{S}^{(PH)}(t)\right)}{\partial x^{(j)}} S^{(PH,i)}(t)$. That is, $\mathcal{P}(\mathbb{x}(t)), \mathbb{x}(t) \in R_+^N, t \geq 0$ satisfies the PDE[51]

(A2.1) $$\sum_{j=1}^{N} \frac{\partial \mathcal{P}(\mathbb{x}(t))}{\partial x^{(j)}} - \mathcal{P}\big(\mathbb{x}(t)\big) = 0, \mathbb{x}(t) \in R_+^N.$$

Setting $\mathcal{P}\big(\mathbb{x}(\mathrm{t})\big) = \mathrm{Q}\big(\mathbb{x}(\mathrm{t})\big) \prod_{\mathrm{k}=1}^{\mathrm{N}} \big(\mathrm{x}^{(\mathrm{k})}\big)^{\gamma^{(\mathrm{k})}}$, $\gamma = \big(\gamma^{(1)}, \dots, \gamma^{(\mathrm{N})}\big) \in \mathrm{R}^{\mathrm{N}}$, it follows that $\mathcal{P}\big(\mathbb{x}(\mathrm{t})\big)$ satisfies (A2.1) if and only if $\sum_{\mathrm{l}=1}^{\mathrm{N}} \gamma^{(\mathrm{l})} = 1$, and

(A2.2) $$\sum_{\mathrm{j}=1}^{\mathrm{N}} \frac{\partial \mathrm{Q}\big(\mathbb{x}(\mathrm{t})\big)}{\partial \mathrm{x}^{(\mathrm{j})}} x^{(\mathrm{j})}(\mathrm{t}) = 0.$$

Next, for all every $\mathrm{a} = \big(\mathrm{a}^{(1)}, \dots, \mathrm{a}^{(\mathrm{N})}\big) \in \mathrm{R}^{\mathrm{N}}, \sum_{\mathrm{k}=1}^{\mathrm{N}} \mathrm{a}^{(\mathrm{k})} = 0$, $\mathrm{Q}(\mathbb{x}) = \sum_{\mathrm{k}=1}^{\mathrm{N}} \mathrm{a}^{(\mathrm{k})} \ln x^{(\mathrm{k})}$ satisfies (A2.2). Thus, the function $\mathcal{P}^{(\mathrm{a},\gamma)}(\mathbb{x}) = \left(\prod_{\mathrm{k}=1}^{\mathrm{N}} \big(x^{(\mathrm{k})}\big)^{\gamma^{(\mathrm{k})}}\right) \big(\sum_{\mathrm{l}=1}^{\mathrm{N}} \mathrm{a}^{(\mathrm{l})} \ln x^{(\mathrm{l})}\big)$ satisfies (A2.1). Thus, all linear combinations of $\mathcal{P}^{(\mathrm{a},\gamma)}(\mathbb{x})$ will solve (A21). In particular, for every $c = \big(c^{(i,j)} > 0, i, j = 1, \dots, N\big), \mathcal{P}^{(c)}(\mathbb{x}) = \sum_{i=1}^{N} \sum_{j=1}^{N} c^{(i,j)} \left(\sqrt{x^{(i)}} - \sqrt{x^{(j)}}\right)^2$ satisfies (A2.1). Furthermore, $\mathcal{P}^{(c)}\left(\mathcal{S}^{(PH)}(0)\right) = 0$ and, for all $t > 0$, $\mathcal{P}^{(c)}\left(\mathcal{S}^{(PH)}(t)\right) = \sum_{i=1}^{N} \sum_{j=1}^{N} c^{(i,j)} \left(\sqrt{S^{(PH,i)}} - \sqrt{S^{(PH,j)}}\right)^2 > 0, \mathbb{P}$-a.s. Thus, $\mathcal{P}^{(c)}\left(\mathcal{S}^{(PH)}(t)\right)$ is an arbitrage portfolio. Q.E.D.

Appendix 2.2: Proof of Proposition 2

[51] Going backward in our arguments shows that (A2.1) is, in fact, a necessary and sufficient condition for the Markov strategy to be self-financing.

Thus, we have the following three equations for the trading strategy $a^{(C,j)}\left(\mathcal{S}^{(PH)}(t)\right), t \geq 0, j = 1, \dots, N$:

$(i)$ $\mathcal{P}^{(\mathrm{C})}\left(\mathcal{S}^{(PH)}(t)\right) = \sum_{j=1}^{N} a^{(C,j)}\left(\mathcal{S}^{(PH)}(s)\right) S^{(PH,i)}(s)$ ;

$(ii)$ $\mathcal{P}^{(\mathrm{C})}\left(\mathcal{S}^{(PH)}(t)\right) - \mathcal{P}\left(\mathcal{S}^{(PH)}(0)\right) = \sum_{j=1}^{N} \int_0^t d^{(S)} a^{(C,j)}\left(\mathcal{S}^{(PH)}(s)\right) S^{(PH,j)}(s) =$

$= \sum_{j=1}^{N} \int_0^t a^{(C,i)}\left(\mathcal{S}^{(PH)}(s)\right) d^{(S)} S^{(PH,j)}(s) + \sum_{j=1}^{N} \int_0^t S^{(PH,j)}(s) d^{(S)} a^{(i)}\left(\mathcal{S}^{(PH)}(s)\right)$;

and

$(iii)$ $\mathcal{P}^{(\mathrm{C})}\left(\mathcal{S}^{(PH)}(t)\right) - \mathcal{P}^{(\mathrm{C})}\left(\mathcal{S}^{(PH)}(0)\right) =$

$$= \sum_{j=1}^{N} \int_0^t \left( a^{(C,i)}\left(\mathcal{S}^{(PH)}(s)\right) - \frac{1}{2} {c^{(j)}}^2(s) \frac{\partial^2 \mathcal{P}^{(\mathrm{C})}\left(\mathcal{S}^{(PH)}(s)\right)}{\partial {x^{(\mathrm{j})}}^2} S^{(PH,j)}(s) \right) d^{(S)} S^{(PH,j)}(s) .$$

Hence,

$$d\mathcal{P}^{(\mathrm{C})}\left(\mathcal{S}^{(PH)}(t)\right) =$$

$$= \sum_{j=1}^{N} a^{(C,i)}\left(\mathcal{S}^{(PH)}(t)\right) - \frac{1}{2} {c^{(j)}}^2(t) \frac{\partial^2 \mathcal{P}^{(\mathrm{C})}\left(\mathcal{S}^{(PH)}(t)\right)}{\partial {x^{(\mathrm{j})}}^2} S^{(PH,j)}(t) d^{(S)} S^{(PH,j)}(t).$$

From, $(ii)$ and $(iii)$, it follows that

$\sum_{j=1}^{N} \left\{ \left( \sum_{i=1}^{N} \frac{\partial a^{(C,i)}\left(\mathcal{S}^{(PH)}(t)\right)}{\partial x^{(j)}} S^{(PH,i)}(t) \right) + \right.$

$\left. \frac{1}{2} {c^{(j)}}^2(t) \frac{\partial^2 \mathcal{P}^{(\mathrm{C})}\left(\mathcal{S}^{(PH)}(t)\right)}{\partial {x^{(\mathrm{j})}}^2} S^{(PH,j)}(t) \right\} S^{(PH,j)}(t) \mu^{(j)}(t) = 0,$

$$\sum_{j=1}^{N}\left\{\left(\sum_{i=1}^{N}\frac{\partial a^{(C,i)}\left(\mathcal{S}^{(PH)}(t)\right)}{\partial x^{(j)}}S^{(PH,i)}(t)\right)+\right.$$

$$\left.\frac{1}{2}{c^{(j)}}^{2}(t)\frac{\partial^{2}\mathcal{P}^{(\mathrm{C})}\left(\mathcal{S}^{(PH)}(s)\right)}{\partial {x^{(\mathrm{j})}}^{2}}S^{(PH,j)}(t)\right\}S^{(PH,j)}(t)\sigma^{(j)}(t)=0$$

From $(i)$, it follows that

$$\mathcal{P}^{(\mathrm{C})}\left(\mathcal{S}^{(PH)}(t)\right)=\sum_{j=1}^{N}a^{(C,j)}\left(\mathcal{S}^{(PH)}(s)\right)S^{(PH,i)}(s)$$

$$d\mathcal{P}^{(\mathrm{C})}\left(\mathcal{S}^{(PH)}(t)\right)=\sum_{j=1}^{N}a^{(C,j)}\left(\mathcal{S}^{(PH)}(s)\right)dS^{(PH,i)}(s)+\sum_{j=1}^{N}S^{(PH,i)}(s)da^{(C,j)}\left(\mathcal{S}^{(PH)}(s)\right)$$

as well as

(A2.3) $$\sum_{j=1}^{N}\left\{\begin{matrix}\frac{\partial\mathcal{P}^{(\mathrm{C})}\left(\mathcal{S}^{(PH)}(t)\right)}{\partial x^{(j)}}-a^{(C,j)}\left(\mathcal{S}^{(PH)}(t)\right)+\\+\frac{1}{2}{c^{(j)}}^{2}(t)\frac{\partial^{2}\mathcal{P}^{(\mathrm{C})}\left(\mathcal{S}^{(PH)}(t)\right)}{\partial {x^{(\mathrm{j})}}^{2}}S^{(PH,j)}(t)\end{matrix}\right\}S^{(PH,j)}(t)\mu^{(j)}(t)=0$$

and

(A2.4) $$\sum_{j=1}^{N}\left\{\begin{matrix}\frac{\partial\mathcal{P}^{(\mathrm{C})}\left(\mathcal{S}^{(PH)}(t)\right)}{\partial x^{(j)}}-a^{(C,j)}\left(\mathcal{S}^{(PH)}(t)\right)+\\+\frac{1}{2}{c^{(j)}}^{2}(s)\frac{\partial^{2}\mathcal{P}^{(\mathrm{C})}\left(\mathcal{S}^{(PH)}(s)\right)}{\partial {x^{(\mathrm{j})}}^{2}}S^{(PH,j)}(t)\end{matrix}\right\}S^{(PH,j)}(t)\sigma^{(j)}(t)=0.$$

Thus, a solution of (A2.3) and (A2.4) is given by

$$a^{(C,j)}\left(\mathcal{S}^{(PH)}(t)\right)=\frac{\partial\mathcal{P}^{(\mathrm{C})}\left(\mathcal{S}^{(PH)}(t)\right)}{\partial x^{(j)}}+\frac{1}{2}{c^{(j)}}^{2}\frac{\partial^{2}\mathcal{P}^{(\mathrm{C})}\left(\mathcal{S}^{(PH)}(t)\right)}{\partial {x^{(\mathrm{j})}}^{2}}S^{(PH,j)}(t),$$

which, together with $(i)$, leads to the PDE

(A2.5) $$\sum_{j=1}^{N}\frac{\partial\mathcal{P}^{(\mathrm{C})}(\mathbb{x})}{\partial x^{(j)}}x^{(j)}-\mathcal{P}^{(\mathrm{S})}(\mathbb{x})+\sum_{j=1}^{N}\frac{1}{2}{c^{(j)}}^{2}\frac{\partial^{2}\mathcal{P}^{(\mathrm{C})}(\mathbb{x})}{\partial {x^{(j)}}^{2}}{x^{(j)}}^{2}=0,\mathbb{x}\in R_{+}^{N}.$$

Then, portfolios of the type $\mathcal{P}^{(C)}(\mathbb{x}) = \prod_{k=1}^{N} \left(x^{(k)}\right)^{\phi^{(k)}}, \phi^{(k)} \in R, k = 1, \dots, N$, satisfy (A2.5) if and only if

(A2.6) $$\sum_{j=1}^{N} \phi^{(j)} - 1 + \sum_{j=1}^{N} \frac{1}{2} c^{(j)^2} \phi^{(j)} \left(\phi^{(j)} - 1\right) = 0.$$

Then the riskless asset $\mathcal{M}$ has dynamics $\beta(t) = e^{rt}$, if $\sum_{j=1}^{N} \sigma^{(j)} \phi^{(j)} = 0$, and $r = \sum_{j=1}^{N} \mu^{(j)} \phi^{(j)}$. Set $S^{(PH,0)}(t) = \beta(t), t \geq 0$. Extend the market $\mathfrak{M}^{(H)}$ to have the price dynamics

(A2.7) $$\frac{\partial \mathcal{P}^{(C)}(t,\mathbb{x})}{\partial t} + r \sum_{j=1}^{N} \frac{\partial \mathcal{P}^{(C)}(t,\mathbb{x})}{\partial x^{(j)}} x^{(j)} - r\mathcal{P}^{(C)}(t,\mathbb{x}) + \sum_{j=1}^{N} \frac{1}{2} c^{(j)^2} \frac{\partial^2 \mathcal{P}^{(C)}(t,\mathbb{x})}{\partial x^{(j)^2}} x^{(j)^2} = 0$$

and set $\mathcal{P}^{(C)}\left(t, x^{(1)}, \dots, x^{(N)}\right) = \mathrm{Q}\left(t, x^{(1)}, \dots, x^{(N)}\right) e^{-bt} \prod_{k=1}^{N} \left(x^{(k)}\right)^{a(k)}$. Then, choosing

$\sum_{j=1}^{N} a(j) = -r - c^{(j)^2}$ and $b = \sum_{j=1}^{N} a(j) - r + \sum_{j=1}^{N} a(j)^2 \frac{1}{2} c^{(j)^2}$, gives

$0 = \frac{\partial \mathrm{Q}\left(t, x^{(1)}, \dots, x^{(N)}\right)}{\partial t} + \sum_{j=1}^{N} \frac{1}{2} c^{(j)^2} \frac{\partial^2 \mathrm{Q}\left(t, x^{(1)}, \dots, x^{(N)}\right)}{\partial x^{(j)^2}} x^{(k)^2}$. Finally, a change of time, $s = T - t$, leading to $\mathrm{V}\left(t, x^{(1)}, \dots, x^{(N)}\right) = \mathrm{Q}\left(T - t, x^{(1)}, \dots, x^{(N)}\right)$, results in

(A2.8) $$\frac{\partial \mathrm{V}\left(t, x^{(1)}, \dots, x^{(N)}\right)}{\partial t} = \sum_{j=1}^{N} \frac{1}{2} c^{(j)^2} \frac{\partial^2 \mathrm{V}\left(t, x^{(1)}, \dots, x^{(N)}\right)}{\partial x^{(j)^2}} x^{(k)^2}$$

The solution of (A2.8) is a multivariate Gaussian density.[52] In particular, if $\mathcal{P}^{(C)}(0, \mathbf{1}) = 0$, then, $\mathrm{Q}(0, \mathbf{1}) = 0$, and thus $\mathrm{V}(T, \mathbf{1}) = 0$, leading to $\mathrm{V}(t, \mathbb{x}) = 0$. That is, no arbitrage strategy is possible. Q.E.D.

---

[52] Consider the multivariate heat equation

Appendix 2.3: Proof of Proposition 3

Without loss of generality, we can assume that $S_H(0) = \beta_H^{(\mathcal{H})}(0) = \beta(0) = 1$. Following the same line of arguments as in the proof of Proposition 1, we obtain the following set of equations:

$(i)$ $\quad \mathcal{P}(x, y, t) = a(x, y, s)x + \beta(t)b(x, y, t) + c(x, y, s)y;$

$(ii)$ $\quad \frac{\partial a(x,y,s)}{\partial x} + \beta(s)\frac{\partial b(x,y,s)}{\partial x} + y\frac{\partial c(x,y,s)}{\partial x} = 0;$

$(iii) - m^{(H)}x^2\frac{\partial a(x,y,s)}{\partial x} - \beta(s)m^{(H)}x\frac{\partial b(x,y,s)}{\partial x} - m^{(H)}xy\frac{\partial c(x,y,s)}{\partial x} +$

$+xyR\frac{\partial a(x,y,s)}{\partial y} + \beta(s)yR\frac{\partial b(x,y,s)}{\partial y} + y^2R\frac{\partial c(x,y,s)}{\partial y} = 0;$

$(iv)$ $\quad \left(x\frac{\partial a(x,y,s)}{\partial t} + \beta(s)\frac{\partial b(x,y,s)}{\partial t} + y\frac{\partial c(x,y,s)}{\partial t}\right) +$

$+\left(\frac{\partial a(x,y,s)}{\partial x}x^2 + \beta(s)\frac{\partial b(x,y,s)}{\partial x}x + xy\frac{\partial c(x,y,s)}{\partial x}\right)\left(m + 2Hm^{(H)}s^{2H-1}\right) +$

$+\left(xy\frac{\partial a(x,y,s)}{\partial y} + \beta(s)y\frac{\partial b(x,y,s)}{\partial y}y + y^2\frac{\partial c(x,y,s)}{\partial y}\right)(v - R2Hs^{2H-1}) = 0.$

---

$$\frac{\partial \mathrm{f}(t, \mathbb{x})}{\partial t} = \frac{1}{2}\sum_{i,j=1}^{N} D_{i,j}\frac{\partial^2 \mathrm{f}(t, \mathbb{x})}{\partial x^{(i)}\partial x^{(j)}}, t \geq 0, \mathbb{x} \in R_+^N,$$

subject to the initial conditions $\mathrm{f}(t, \mathbb{x}) = \delta(\mathbb{x})$, where $D = [D_{i,j}] = D^T$ is constant matrix of diffusion constants. Then,

$$\mathrm{f}(t, \mathbb{x}) = \frac{1}{(2\pi t)^{\frac{N}{2}}\sqrt{|detD|}}\exp\left\{-\frac{1}{2t}\ \mathbb{x}^{\mathrm{T}}D^{-1}\ \mathbb{x}\right\}, t \geq 0, \mathbb{x} \in R_+^N,$$

see, for example, Chirikjian (2009, Chapter 2, Section 2.6.2).

The solution of the system of equation satisfies the following heat equation with zero force term:

$$-r\mathcal{P}(x,y,t) + rx\frac{\partial\mathcal{P}(x,y,t)}{\partial x} + ry\frac{\partial\mathcal{P}(x,y,t)}{\partial y} + \frac{\partial\mathcal{P}(x,y,t)}{\partial t} = 0.$$

Consider $\mathcal{P}(x,y,t) = \left(\sqrt{x} + \sqrt{y} - 2e^{\frac{r}{2}t}\right)^2$. Then $\mathcal{P}(x,y,t)$ determines a self-financing strategy and, furthermore, $\mathbb{P}(\mathcal{P}\left(S_H(0), \beta_H^{(\mathcal{H})}(0), 0\right) = 0,\ \mathcal{P}\left(S_H(T), \beta_H^{(\mathcal{H})}(T), T) \geq 0\right) = 1.$ Q.E.D.

## REFERENCES

Aurzada F. and Mönch C. (2016) Persistence probabilities and a decorrelation inequality for the Rosenblatt process and Hermite processes, *arXiv:1606.00236[*math.PR], https://arxiv.org/pdf/1607.04045v1.pdf

Bai S., Taqqu M.S. (2014) Generalized Hermite processes, discrete chaos and limit theorems, *Stochastic Processes and their Applications*, 124, 1710–1739.

Bender C, Sottinen T, Valkeila E. (2007) Arbitrage with fractional Brownian motion? *Theory of Stochastic Process*, 13, 23–34.

Basse-O'Connor A. and Weber M. (2016) On the Φ-variation of stochastic processes with exponential moments, *Transactions of the London Mathematical Society*, 3, 1-27.

Biagini F., Hu Y., Øksendal B., Zhang T. (2008) *Stochastic Calculus for Fractional Brownian Motion and Applications*, Springer, London.

Bender C., Sottinen T. and Valkeila E. (2011) Fractional processes as models in stochastic finance, in G. Di Nunno, B. Øksendal (eds.), *Advanced Mathematical Methods for Finance,* Springer, Berlin

Black, F., 1972, Capital market equilibrium with restricted borrowing. *Journal of Business* 45, 444-455.

Black, F., and Scholes M. (1973) The pricing of options and corporate liabilities, *Journal of Political Economy* 81, 637-654.

Björk T., and Hult H. (2005) A note on Wick products and the fractional Black–Scholes model. *Finance and Stochastics,* 9, 197–209.

Chen Z., Xu L. and Zhu D. (2015) Generalized continuous time random walks and Hermite processes, *Statistics & Probability Letters*, 99, 44–53.

Cheridito, P. (2001) Mixed fractional Brownian motion. *Bernoulli*, 7, 913-934.

Cheridito, P. (2003) Arbitrage in fractional Brownian motion models. *Finance and Stochastics*, 7, 533-553.

Chirikjian G.S. (2009) *Stochastic Models, Information Theory, and Lie Groups, Vol. 1*. Classical Results and Geometric Methods, Birkhäuser, Boston.

Chronopoulou A. (2008) Self-similarity/memory-length parameter estimation for non-Gaussian Hermite processes via chaos expansions, in Skiadas C.H., Dimotikalis I. and Skiadas C. (ed.) *Topics on Chaotic Systems,* World Scientific New Jersey, 79-86.

Chronopoulou A., Tudor C.A., Viens F.G. (2011) Self-similarity parameter estimation and reproduction property for non-Gaussian Hermite processes, *Communications on Stochastic Analysis*, 5, 161-185.

Clausel M., Roueff F., Taqqu M.S., C. Tudor C. (2014) Asymptotic behavior of the quadratic variation of the sum of two Hermite processes of consecutive orders, *Stochastic Processes and their Applications*, 124, 2517–2541.

Dehling H., and Taqqu M.S. (1989) The empirical process of some long-range dependent sequences with an application to U-statistics, *The Annals of Statistics*, 17, 1767-1783.

Dobrushin R.L. (1979) Gaussian and their subordinated self-similar random generalized fields, *Annals of Probab*ility, 7, 1–28.

Dobrushin, R.L., Major P. (1979) Non-central limit theorems for non-linear functional of Gaussian fields, *Probability Theory and Related Fields*, 50, 27–52.

Duncan, T.E., Hu, Y., Pasik-Duncan,B. (2000) Stochastic calculus for fractional Brownian motion, *SIAM Journal Control Optimization*, 38, 582-612.

Duffie D. (2001) *Dynamic Asset Pricing Theory*, Princeton University Press, Princeton, Oxford

Elliott RJ, Van der Hoek J. (2003) A general fractional white noise theory and applications to finance, *Mathematical Finance*, 13, 301–330.

Embrechts P. and Maejima M. (2002) *Selfsimilar Processes*, Princeton University Press, Princeton, New Jersey.

Fauth A. and Tudor C.A. (2016) Multifractal random walk driven by a Hermite process, in Florescu, I., Mariani, M.C., and Stanley H.E. (edt.) *Handbook of High-Frequency Trading and Modeling in Finance*, Chapter 8, 221-249.

Fisher P.R. (2013) Reflections on the meaning of "risk free", in *Sovereign risk: a world without risk-free assets?* BIC Papers No 72, July 2013.

Föllmer, H. and Schied A. (2013) Probabilistic aspects of finance, *Bernoulli*, 19(4), 1306-1326.

Guasoni P, Rásonyi, M., and Schachermayer W. (2008) Consistent price systems and face-lifting pricing under transaction costs, *The Annals of Applied Probability*, 18, 491-520.

Gu H., Liang, L.R., and Zhang, Y.X., (2012) Time-changed geometric fractional Brownian motion and option pricing with transaction costs, *Physica A*, 391, 3971-3977.

Guasoni, P., Rásonyi, M., and Schachermayer. W., (2008) Consistent price systems and face-lifting pricing under transaction costs, *The Annals of Applied Probability*, 18, 491-520.

Hu Y., Øksendal B. Fractional white noise calculus and applications to finance. *Infinite Dimensional Analysis, Quantum Probability and Related Topics*, 6, 1–32.

Kozachenko Y., Melnikov A., and Mishura Y. (2014) On drift parameter estimation in models with fractional Brownian motion, *Statistics: A Journal of Theoretical and Applied Statistics*, 49, 35-62.

Kuznetsov, Yu.A. (1999) The absence of arbitrage in a model with fractal Brownian motion, *Russian Mathematical Surveys*, 54,847-848.

Lacey M.T. (1991) On weak convergence in dynamical systems to self-similar processes with spectral representation, *Transactions of the American Mathematical Society*, 328, 767-778.

Lavancier F.(2006) Long memory random fields, in Bertail, P., Doukhan, P., Soulier, P. (Eds.) *Dependence in Probability and Statistics*, Springer, 195-220 .

Maejima M., Tudor C.A. (2007) Wiener integrals with respect to the Hermite process and a non-central limit theorem, *Stochastic Analysis and Applications*, 25, 1-13.

Marty R. (2013) From Hermite polynomials to multifractional processes, *Journal of Applied Probability*, 50, 323-343.

Merton, R. (1973) Theory of rational option pricing, *Bell Journal of Economics and Management Science* 4, 141–183.

Mishura Y. (2008) *Stochastic Calculus for Fractional Brownian Motion and Related Processes*. Springer, Berlin.

Mishura Y. S. and Valkeida E. (2002) The absence of arbitrage in a Mixed Brownian-Fractional Brownian model. *Proceedings of the Steklov Institute of Mathematics*, 237, 215-224

Molchan, G. (1969) Gaussian processes with spectra which are asymptotically equivalent to a power of λ, *Theory of Probability and its Applications*, 14, 530–532

Molchan, G., Golosov, J., (1969) Gaussian Stationary processes with asymptotic power spectrum, *Soviet. Math. Doklady,* 10, 134-137.

Necula. C., 2002. Option pricing in a fractional Brownian motion environment, *Academy of Economic Studies Bucharest*, *Romania,* Preprint, Academy of Economic Studies, Bucharest.

Neuenkirch A. and Nourdin I. (2007) Exact rate of convergence of some approximation schemes associated to SDEs driven by a fractional Brownian motion, *Journal of Theoretical Probability*, 20, 871-899.

Nteumagné, B.F., Pindza, E. , and Maré E. (2014) Applying the barycentric Jacobi spectral method to price options with transaction costs in a fractional Black-Scholes framework, *Journal of Mathematical Finance,* 2014, 4, 35-46.

Nualart D. (2006) *The Malliavin Calculus and Related Topics*, Springer, Berlin.

Pipiras, V., Taqqu M.S. (2010) Regularization and integral representations of Hermite processes, *Statististics & Probability. Lett*ers, 80, 2014–2023.

Poularikas A. D. (1999). Hermite polynomials, in *The Handbook of Formulas and Tables for Signal Processing*, A. D. Poularikas (ed.) Boca Raton: CRC Press LLC, Chapter 22.

Rachev S.T, and Fabozzi F.J. (2016) Financial markets with no riskless (safe) asset, *arXiv:1612.02112 [q-fin.MF],* https://arxiv.org/abs/1612.02112.

Rogers, L.C.G. (1997) Arbitrage with fractional Brownian motion, *Mathematical Finance,* 7, 95–105.

Rostek S. (2009) Option *Pricing in Fractional Brownian Markets*, Lecture Notes in Economics and Mathematical Systems 622, Springer, Berlin.

Russo F., and Vallois P. (1993) Forward, backward and symmetric stochastic integration, *Probability Theory and Related Fields*, 97, 403-421.

Samorodnitsky, G. (2016) *Stochastic Processes and Long Range Dependence*, Springer, New York.

Seneta E. (1976) *Regularly Varying Functions*, Lecture Notes in Mathematics, 508, Springer, Berlin.

Shiryayev, A.N. (1998) On arbitrage and replication for fractal models, Research Report 20, *MaPhySto*, Department of Mathematical Sciences, University of Aarhus, Denmark.

Shokrollahi, F. and Kılıçman A. (2016) The valuation of currency options by fractional Brownian motion, *SpringerPlus DOI: 10.1186/s40064-016-2784-2.*

Sottinen, T. (2001) Fractional Brownian motion, random walks and binary market models, *Finance and Stochastics* 5, 343–355.

Sottinen, T., and Valkeila E. (2003) On arbitrage and replication in the fractional Black–Scholes pricing model, *Statistics & Decisions*, 21, 137–151.

Sun X., and Cheng R. (2014) A weak convergence to Hermite process by martingale differences, *Advances in Mathematical Physics*, 2014, Article ID 307819, 10 pages

Taqqu, M.S. (1979) Convergence of integrated processes of arbitrary Hermite rank, *Zeitschrift für Wahrscheinlichkeitstheorie und Verwandte Gebiete*, 50, 53–83

Taqqu M.S. (2011) The Rosenblatt process, in *Selected Works of Murray Rosenblatt*, R.A. Davis et al. (eds) *Selected Works in Probability and Statistics*, Springer Science + Business Media, New York, 29-45.

Torres S., Tudor C.A (2009) Donsker type theorem for the Rosenblatt process and a binary market model, *Stochastic Analysis and Applications*, 27, 555-573.

Tsoi A. (2011) Fractional white noise multiplication, in Tsoi, A. Nualard D., and Yin G. (eds) *Stochastic Analysis, Stochastic Systems, and Applications to Finance*, World Scientific, New Jersey, 2011, 27-42.

Tudor C.A. (2008) Analysis of the Rosenblatt process, *ESAIM: Probability & Statistics,* 12, 230–257.

Tudor C.A. (2013) *Analysis of Variations for Self-similar Processes: A Stochastic Calculus Approach*, Springer Probability and Statistics Series, Springer, New York.

Wang X-T. (2010) Scaling and long-range dependence in option pricing I: Pricing European option with transaction costs under the fractional Black–Scholes model, *Physica A: Statistical Mechanics and its Applications*, 389, 438–444

Xiao W-L., Zhang W-G., Zhang X-L., and Wang Y-L. (2010) Pricing currency options in a fractional Brownian motion with jumps, *Economic Modelling* 27, 935–942.

Zähle, M. (1998) Integration with respect to fractal functions and stochastic calculus, I *Probability Theory and Related Fields*, 111, 333-374.

Zähle, M. (2002) Integration with respect to fractal functions and stochastic calculus. *Math. Machr.II*, 225, 265-280.